# An Interpretable Machine Learning Approach in Predicting Inflation Using Payments System Data: A Case Study of Indonesia[1]

**Wishnu Badrawani** [a†]
[a] Bank Indonesia Institute, Bank Indonesia, Jakarta, Indonesia.

## Abstract:

This paper evaluates the performance of prominent machine learning (ML) algorithms in predicting Indonesia's inflation using the payment system, capital market, and macroeconomic data. We compare the forecasting performance of each ML model, namely shrinkage regression, ensemble learning, and super vector regression, to that of the univariate time series ARIMA and SARIMA models. We examine various out-of-bag sample periods in each ML model to determine the appropriate data-splitting ratios for the regression case study. This study indicates that all ML models produced lower RMSEs and reduced average forecast errors by 45.16 percent relative to the ARIMA benchmark, with the Extreme Gradient Boosting model outperforming other ML models and the benchmark. Using the Shapley value, we discovered that numerous payment system variables significantly predict inflation. We explore the ML forecast using local Shapley decomposition and show the relationship between the explanatory variables and inflation for interpretation. The interpretation of the ML forecast highlights some significant findings and offers insightful recommendations, enhancing previous economic research that uses a more established econometric method. Our findings advocate ML models as supplementary tools for the central bank to predict inflation and support monetary policy.

---

[1] The article is part of a PhD thesis of the author at the University of Birmingham

† Corresponding author. Email address: w_badrawani@bi.go.id

## INTRODUCTION

Understanding the current economy is essential for policymakers to make sound decisions. The conventional forecasting technique is reliable under normal circumstances; however, macroeconomic forecasting during times of crisis and extraordinary circumstances, such as the COVID-19 pandemic, has been empirically proven to be challenging. According to Juhro and Syarifuddin (2017), prior to the pandemic crisis, the global economy was already fragile, uncertain, complex, and ambiguous. Thus, forecasting economic indicators using traditional linear regression approaches may become tenuous due to their limited ability to identify complex variables with abundant data and information. As a result, policymakers have begun to consider alternatives to the time-series regression models traditionally employed for forecasting critical macroeconomic variables, such as Artificial Intelligence (AI) and Machine Learning (ML) approaches, so they can closely monitor the economy, see Doerr et al. (2021).

ML has many advantages despite concerns regarding causal inference, result interpretation, and determining which factors should be considered when making a specific policy recommendation. Its adaptability allows it to uncover complex hidden structures that need to be explicitly stated and accommodate massive data sets with many regressors without overfitting (Mullainathan & Spiess, 2017). Additionally, Athey and Imbens (2019) suggest that researchers may untangle more sophisticated and multidisciplinary research by exploring these approaches.

The Indonesian economy has effectively maintained low inflation in recent years, notably since the central bank implemented inflation targeting. However, it is critical to provide the most accurate and timely inflation rate forecast for emerging economies, particularly in exceptional circumstances such as during economic crises or pandemics. Considering the interpretability limitation of the ML approach, making ML prediction more accessible and understandable would be advantageous and provide more meaningful insight into policy recommendations. In addition, Aprigliano et al. (2019) argue that payment system data can be employed as an alternative data source for macroeconomic forecasting due to its characteristic of recording all economic transactions between various entities. Therefore, this paper aims to investigate the use of ML models with an interpretability approach that explores payment system data, capital market, and macroeconomic data to estimate inflation rates in Indonesia.

Accordingly, the current study contributes to the body of knowledge in various ways. First, this study contributes to a growing body of literature that successfully compares ML models to the established econometric techniques in economic forecasting, incorporating a novel approach to the interpretability of ML forecasts. Second, our study compares the ML model performances in different data-splitting ratios, a topic that needs to be adequately covered. Lastly, to the best of our knowledge, this is the first study to use interpretable machine-learning techniques to forecast inflation using payment system data in Southeast Asian countries.

The remainder of the essay is organized as follows. Section two discusses previous studies. Sections three and four describe this study's dataset and methodology. Section five discusses all observed ML models' empirical outcomes, and Section six summarises the study's conclusions. The appendices of this study include a brief description of payment system data and other observable variables.

## LITERATURE REVIEW

### Machine learning and economic forecasting

Machine learning (ML) is commonly used for prediction, processing vast amounts of data by clustering and grouping, and providing a classification with an advanced algorithm in a 'black box' for automation purposes. Nevertheless, ML functionality, mainly supervised ML, has rapidly advanced into the field of interpretation and causal inference and has been effectively implemented in practice (Athey & Imbens, 2019).

Many studies have examined various ML methodologies to forecast macroeconomic variables such as economic growth, and the number of studies in the literature is rising. Yoon (2021) uses ensemble learning approaches such as Gradient Boosting and the Random Forest algorithm to predict GDP growth in Japan, whereas Milacic et al. (2017) and Lepetyuk et al. (2020) employ neural networks and deep learning methods using the US dataset. Several researchers have also compared various ML methods to forecast macroeconomic variables; for example, see Richardson et al. (2021) and Tamara et al. (2020). Their result suggests that many ML methods outperformed the traditional econometric approach. Numerous research in banking and finance domains have also found that the use of ML models for forecasting considerably increases the accuracy of their results. Heaton et al. (2017) and Zhang and Chen (2021), for instance, employ deep learning models for financial forecasting and solve stock market index problems.

In addition to employing financial and economic data for economic forecasting, alternative data from payment systems have been utilised in quantitative research to forecast macroeconomic indicators, especially in developed countries. For instance, Galbraith and Tkacz (2018) showed that payment-system data, particularly debit card and check transactions, can improve GDP forecasting, which is supported by Chapman and Desai (2020). In support of these findings, Barnett et al. (2021) have criticised the theory that suggests simple sum monetary indicators, i.e., ignoring credit cards from money supply measurement, are incompatible with the economic theory that develops the models using various data.

Several researchers have used ML to help optimise central bank policy. Chakraborty and Joseph (2017), for example, offered several ML models for central bank usage and policy context, whereas Zulen et al. (2018) employed text analysis to determine stakeholders' expectations to predict the central bank's policy rate. Various ML algorithms have also been applied to forecast inflation rate and suggested to optimise central bank policy; for example, see Medeiros et al. (2021) and Rodriguez-Vargas (2020). Lu (2020) examines financial market risk indicators that can promote monetary policy effectiveness through macro-control monetary policy adoption. In recent years, interpretable ML has gained popularity by extending the focus of ML from optimising forecast accuracy (or minimising forecast error) to the interpretability of forecasting results; for example, see Molnar (2020). Interpretability approaches are not limited to measuring the importance of input variables for model prediction but also attempt to reveal the causal inference approach of ML models in macroeconomic forecasting (Buckmann et al., 2021).

In the context of Indonesia, there is a limited study that employs ML methods to examine macroeconomics, particularly inflation. Some authors have used the long-short term memory (LSTM) method to forecast the inflation rate (Zahara & Ilmiddaviq, 2020, Savitri et al., 2021). These papers are like mostly ML studies in regression that emphasise maximising prediction accuracy. However, there are no causal inferences of the observed variables in the model, and it is unclear how the ML prediction result can be helpful for policy analysis. According to Banerjee et al. (2005), there is evidence that simple univariate forecasting algorithms are difficult to beat and often more accurate than more complicated multivariate models; especially for short-term period forecasting (Chayama & Hirata, 2016).

Many previous studies examining the performance of ML models to predict macroeconomic indicators used univariate econometric models such as AR(Auto-Regressive) or ARIMA (Auto-Regressive-Integrated-Moving-Average) models as a benchmark (Richardson et al., 2021, Ozgur & Akkoc, 2021). Therefore, this study employs ARIMA as the benchmark model to examine the performance of various ML models. In addition, the absence of studies on inflation using non-traditional explanatory variables in developing countries, particularly Indonesia, encourages us to examine non-traditional data for inflation forecasting, such as payment systems data and a few data from the capital market.

**Literature Related to Explanatory Variables**

According to Bank Indonesia (2019), a payment system is a legal framework composed of institutions and procedures for the transfer of funds to fulfil liabilities emerging from economic activities. Commonly, the payment system is classified into two categories based on transaction size, high-value payment systems that served by RTGS (Real-Time Gross Settlement) and retail payment systems that served by the national clearing system, the fast payment, and card or electronic money platform. Theoretically, the advancement of payment instruments has a reversal effect on inflation by reducing the money demand for transactions in the economy (McCallum & Goodfriend, 1989), particularly short-term inflation (Csonto et al., 2019). Lubis et al. (2019) and Adil et al. (2020) found that technological advancement in the payment system negatively impacts the demand for money as the intermediary target of monetary policy in controlling inflation.

Some economic literature discusses the negative effect of inflation on the capital market; for example, a higher rate of inflation can cause a substantial reduction in the ratio of share prices to tax earnings (Albulescu et al., 2017), and its effect significantly occurs when monetary policy is countercyclical (Zhang, 2021). Juhro et al. (2020) found that stock market capitalisation was a negative and significant predictor of inflation, which corroborates Wahyudi et al. (2017) for South Asia countries except for Thailand. Inflation also has a negative effect on short-term stock market returns but a positive effect in the long term (Hashmi et al., 2021).

The exchange rate, in theory, can affect inflation directly through imported consumer goods and indirectly through imported intermediate goods. The exchange rate's effect on inflation is the highest and most immediate for import prices, and the impact decreases along the pricing chain (Ortega & Osbat, 2020), which exchange rates and inflation fluctuations may also result from monetary policy (Gortz et al., 2020). According to Warjiyo and Juhro (2019), in a small open economy like Indonesia, exchange rate movements may not necessarily represent its fundamental value, and depreciation of the rupiah had no significant pass-through effects on tradeable core inflation.

According to Maehle (2020) and Juhro et al. (2020), the interest rate has a negative effect on inflation in the monetary policy framework by impacting several channels and intermediary targets. The effect of the policy rate on inflation is not immediate and displays distributed lags (Warjiyo & Juhro, 2022). Lukmanova and Rabitsch (2020) highlighted that a temporary nominal interest rate shock causes inflation and economic activity to fall, while a persistent inflation target increase causes the nominal interest rate, inflation, and economic activity to rise. An immediate reduction in Bank Indonesia's inflation target may lead to a decrease in average inflation and nominal interest rates, and vice versa (Lie, 2019). One argument is that policymakers may have reduced the monetary policy variable to anticipate future inflation; as a result, prices rise, albeit lower than they would have if the policy rate had not been raised. Consequently, the causal interpretation of inflation and interest rate may differ with respect to the role of inflation expectation, expected rate, channel utilised, and level uncertainty.

## DATASET

This paper aims to examine the relationship between payment system variables and inflation employing ML and econometric models. We observe various payment system variables that appear to have a close relationship with the inflation rate, including retail payment instruments that track transactions between individuals and small business entities and wholesale payments that could represent the entire economic activity.

We identify the variables in the payments system, macroeconomics, and capital market data that are available for analysis, aggregated into monthly data from January 2015 to December 2021, as depicted in Figure 1. The dependent variable, inflation, is the monthly percentage change in the consumer price index (year on year) based on the year 2012, as shown in Figure 1a. In addition to payment system data and macroeconomic factors, we also incorporated capital market statistics, as depicted in Figures 1b and 1c. The variables used in our dataset are listed and described in detail in Appendix A. All data were collected from Bank Indonesia and the Indonesian Bureau of Statistics.

**Insert Figure 1 (sub figure a, b, c)**

## METHODOLOGY

This section briefly describes the various ML and benchmark models we considered for forecasting inflation; however, see the original literature for greater detail. In classic econometrics, particularly linear regression, one of the objectives was to infer a causal relationship between regressors and target variable; however, high correlations between features can result in multicollinearity issues, resulting in interpretive bias (Brooks, 2019). ML techniques, on the other hand, emphasise predictive accuracy and minimising forecast error in regression problems (Ajiga et al., 2024).

This research applies some of the acknowledged ML models, including the ridge, lasso regression, and elastic net. These models are well-known for using regularisation techniques that employ shrinkage estimators to lower the prediction variance by reducing the parameter estimates in the standard OLS model. Thus, it truncates model complexity when adding many variables (Smeekes & Wijler, 2018). This work also employs other ML techniques, like support vector machines and ensemble learning, which are less susceptible to multicollinearities (Chan et al., 2022). All the observed ML models are supervised models, where the dataset is labelled and defined to train the algorithm in making a prediction.

## Benchmark and Machine Learning Models

### *Benchmark Models and SARIMA*

The *ARIMA* model is used as a benchmark in this study to evaluate the forecast performance of the observed ML methods. The regular ARIMA model incorporates the autoregressive and moving average terms, and the seasonal ARIMA model *(SARIMA)* was developed by incorporating seasonal factors into the ARIMA model.

### *Ordinary Least Square (OLS) and Shrinkage Regression*

The main objective of an estimation technique is to fit the target variable's prediction function of equation $y = f(X) + e_t$ using potential regressors to determine the target variable's forecasted values, where *y* denotes the dependent variable, *f* is some fixed but unknown function of regressors $X_1, X_2, \ldots, X_n$, and $e_t$ is a random error term. However, under penalty function constraints, shrinkage estimators in shrinkage models (such as the ridge, lasso, and elastic net) reduce the least square errors to the absolute minimum.

*Ridge regression* is a method for estimating the coefficients of multiple regression that is similar to least squares but uses a penalty in the form of L2 regularisation, where the penalty is on the squared magnitude of the coefficients, to minimise the coefficient of each regressor (Hastie et al., 2017).

*Lasso regression* is a shrinkage-based alternative to ridge regression that overcomes the difficulty of omitting irrelevant features from the regression equation, using L1 regularisation to create more parsimonious models by imposing a penalty with a value equal to the absolute value of the coefficients' magnitude (James et al., 2013). The L1 penalty function has the effect of forcing some of the coefficient estimates to be exactly equal to zero; thus, the lasso narrows the coefficient estimates towards zero during the regularisation process and can be employed for feature selection.

*The Elastic net regression* is a hybrid of lasso and ridge regression methods that overcomes the shortcomings of lasso and ridge algorithms when dealing with highly correlated data. The combination of L1 and L2 penalty functions allows for learning the sparse selection while stabilising regularisation paths and removing limitations on the number of selected variables.

### *Ensemble Learning Model*

*Random forest (RF)* is a decision tree–based ensemble learning method built using a forest of many regression trees that reduces the variance error of a model's result. It is a non-parametric method; thus, it solves the multicollinearity problem slightly different than parametric approaches such as OLS or lasso regression. It uses a bagging (bootstrap aggregation) approach, i.e., by introducing randomness from a subset of the training dataset where each tree is independently built. The output of trees is then averaged to produce predictions (Muller & Guido, 2016). By averaging over numerous trees, volatility is reduced, and predictive performance is stabilised. Because each tree generated during the bagging process is identically distributed (i.d.), thereby lowering variance and improving forecast performance (Hastie et al., 2017).

*Extreme Gradient Boosting (XGB)* is a scalable boosting ensemble learning developed by Chen and Guestrin (2016). The XGB model integrates several weak learner trees to develop a strong learner through additive learning, includes regularisation to prevent overfitting and improves the training process. It was improved based on a gradient-boosting algorithm.

### *Support Vector Regression (SVR)*

*Support Vector Regression (SVR)* is one of the popular supervised ML algorithms, which can be used for both classification and regression problems. It uses a distinct objective function compared to the least-squares or the shrinkage methods such as ridge, lasso, and elastic net regression. Hamel (2011) iterates that the SVM algorithm tries to find a line/hyperplane (in multidimensional space) that predicts or separates a data set into different classes, for example, two different groups, by using a small subset of training data points (called support vectors).

## Model Estimation and Hyperparameter Optimisation

The empirical analysis begins with partitioning the dataset into test and training datasets for the regressor and the target variable. The rule of thumb for the data splitting ratio commonly used in the classification case study is 70:30 (Nguyen et al., 2021; Vrigazova, 2021). To our knowledge, there is no benchmark or guideline for data splitting ratio in the regression study case of ML models. Due to the small size of the dataset, we used the first 80 percent of observations as a training set, while the remaining 20 percent of the data were used to evaluate the out-of-sample predicting accuracy (80:20). The testing dataset will be extended to various ranges from 6 months, nine months, 12 months and a maximum period of 24 months (70:30), as commonly used by several central banks for inflation prediction, to determine the most effective data-splitting ratio with the lowest forecast error.

We employed a cross-validation technique to avoid overfitting (Figure 2); the testing subset set aside during the training process was then used to examine the model's generalisability (out-of-sample forecasting). We employ Python 3 with various libraries such as NumPy, Pandas, Sklearn, Matplotlib, and Statsmodels to perform analysis and predictions. In this arrangement, the model is then estimated using data from the training set. Finding the parameters that minimise the forecast error in the training data set is essential for selecting the optimal hyperparameters. Once the hyperparameter tuning is complete, we will use the test dataset to evaluate the fitted model's out-of-sample predictive accuracy. GridSearch is used to cross-validate hyperparameters to determine the best ones. See Feurer and Hutter (2019) and Yang and Shami (2020) for more details. Based on the parameters suggested by GridSearch, we find the best hyperparameter with the lower MSE and RMSE.

---

**Insert Figure 2**

---

This procedure is carried out without involving the test dataset. To avoid overfitting the model to the training dataset, we restrict our search for hyperparameter values to a range of values based on prior research. This procedure enables the fitted models to analyse the unobserved test dataset. Table 1 summarises the hyperparameters optimised for each model, the range of values explored, and the proposed optimal values. When no value is supplied for a parameter, we use the default value of the Python ML algorithm.

---

**Insert Table 1**

---

**Forecast Evaluation Methodology**

After applying the obtained ML models to the test dataset, all proposed models are evaluated by comparing their forecast accuracy to the benchmark using various parameters such as RMSE (root mean square error) and MAE (mean absolute errors). MAE offers us an indication of the magnitude of the mistake and is a measure strongly preferred and frequently used by both practitioners and academics to assess the accuracy of ML forecasts (Das & Cakma, 2018). We also measure the forecast accuracy of each model by calculating the RMSE, the average sum of squared errors. The lower the RMSE score, the better. Additionally, the Diebold–Mariano (DM) statistic test (Diebold & Mariano, 2002) is used to see if each ML model forecast differs considerably from the benchmark model. By confirming the existence of the null hypothesis, the DM test determines the statistical significance by comparing the averages of the two loss functions (we choose MSE), thereby indicating that both models provide the same level of forecast accuracy.

**Feature Importance and Interpretation**

In this study, we examine the feature importance of the proposed ML models using SHAP values, as suggested by Lundberg et al. (2020). Shapley's feature importance is model agnostic, which means that it can be applied to ML model without regard to its specific characteristics.

Shapley values are a measure of marginal contributions from a given feature on a particular model. This method originated from cooperative game theory and was first introduced by Lundberg and Lee (2017) for the interpretation of ML prediction. It provides a general solution to the problem of attributing a reward obtained in a cooperative game to the individual players based on their contribution to the game, which is called SHAP (SHapley Additive exPlanations). Shapley values $\phi$ measure feature importances based on the relative contribution of a feature $i$, weighted and summed across all possible feature value combinations and is formally defined via a value function $v$ given by:

$$\phi_{i(v)} = \sum_{S \subseteq F \setminus \{i\}} \frac{|S|!(|F|-|S|-1)!}{|F|!} \left[f_{S\cup\{i\}}\left(x_{S\cup\{i\}}\right) - f_S(x_S)\right] \tag{1}$$

where $S$ is a subset of the features used in the model and $F$ is the complete set of all features, $f_{S\cup\{i\}}$ is a trained model that includes a given feature present, and $f_S$ is a trained model that does not include the appointed feature. Then, we compare predictions from the two models on the current input $f_{S\cup\{i\}}\left(x_{S\cup\{i\}}\right) - f_S(x_S)$, where $x_S$ represents the values of the input features in the set $S$. Following the computation of the preceding differences for all feasible subsets $S \subseteq F\setminus\{i\}$, the Shapley values are calculated and applied as feature importances that are averagely weighted over all conceivable differences.

SHAP values are used to describe the change in model prediction when a feature is included in the model. Given a model $f(z)$, SHAP explains how features combine additively to shift the model's conditional expectation $E[f(z)|z = x]$ away from the base value $E[f(z)]$. Then we average all these differences and result in:

$$\phi_i(x) = \frac{1}{M}\sum_{m=1}^{M}\phi_i^m \quad (2)$$

Obtaining aggregate Shapley values requires the technique to be performed repeatedly for each feature, where $M$ refers to the number of iterations; for a more detailed explanation, see Molnar (2020). Although the aggregate analysis shows which variables are important, it does not explain how the model learned its functional form. Following Buckmann et al. (2022) and Lundberg et al. (2019), a local Shapley decomposition will be carried out to investigate the importance of local factors in an individual prediction for each feature, based on the fitted data.

## EMPIRICAL RESULTS

### Model Forecast Performance

This section presents the result of the inflation forecast obtained by various ML algorithms compared to a time series model based on payment system data from January 2015 to December 2021 with a training-testing data split ratio of 80:20. The results found that all ML models generate forecasts with lower RMSE than the benchmark. Indicated by the lowest RMSE and MAE values and DM test statistics, the XGB model outperforms univariate regression and other ML models can reduce forecast errors by 53.22% relative to the benchmark.

**Insert Table 2**

The DM test statistics and the p-values of the DM test statistics, which are shown in the fourth and fifth columns of Table 2, are employed to evaluate the significance of forecast accuracy for each model relative to the benchmark. The results of the DM test indicate that the null hypothesis can be rejected, indicating that there are statistically significant differences between the forecasts produced by the ML models and the ARIMA benchmark. The results of the MAE and RMSE indicate that the ML method significantly improves the average forecast accuracy of the ARIMA benchmark's inflation rate prediction by 45.16%.

Figure 3 depicts the actual inflation rate and out-of-sample prediction derived from each ML model from September 2020 to December 2021. It is evident that most ML models appropriately predict inflation volatility, with the XGB model exhibiting the best performance, followed by the lasso regression.

**Insert Figure 3**

The ensemble learning model, notably XGB regression, produces exceptional outcomes in the out-of-sample prediction, while RF does not perform well. Comparing the prediction outcomes for the entire sample period (training and testing datasets) of all ML models, we discovered that XGB regression models had the lowest RMSE and MAE. In addition to achieving the lowest forecast errors, the XGB was also able to precisely estimate the inflation rate in the case of an inflation decline in 2016 and throughout the COVID-19 pandemic period beginning in 2020 (Figure 4). Furthermore, the penalised regression model, lasso regression, appropriately predict the inflation rate for the next sixteen months. This study corroborates Gosiewska et al. (2021) assertion that a simple ML model can achieve comparable performance to a complicated ML model, which is advantageous for interpretability and policy recommendation.

**Insert Figure 4**

Our findings suggest that the XGB regression outperforms alternative ML and benchmark model, which supports earlier empirical studies. For example, Yoon (2021) and Richardson et al. (2021) forecast GDP using a macroeconomic dataset, and Lundberg et al. (2019) predict mortality risk and kidney diseases with some interpretation.

### Impact of training and testing data split on accuracy

In addition to evaluating the model forecast performance, the proposed models were also fitted to various data split ratios to evaluate their forecast performance in different scenarios. Figure 5 depicts the plotted values of the RMSE for inflation rate prediction from time series and various ML models with varying data partitioning.

---

**Insert Figure 5**

---

The *x*-axis of Figure 5 depicts the various models examined and the *y*-axis indicates the RMSE for inflation rate forecasts. We analyse the forecast error of different models with varying data splitting ratios and months (24, 16, 12, 9, and 6 months). In general, the RMSE of the model's predictions decreases as the period of the testing data decreases and the period of the training data set increases. However, depending on the size of the training and testing sets, the accuracy of each algorithm may vary. Despite ensemble learning's remarkable and consistent prediction ability, RF and XGB regression show a substantial increase in predicting error over the longer term (24 months).

Although most ML models can produce MAEs and RMSEs that are less than those of the benchmark in various out-of-sample periods, the prediction performance of univariate time series regression, ARIMA and SARIMA, on 12-month test data was found to be superior to that of all ML models except for XGB. This finding confirms Banerjee et al. (2005), who found that multivariate models did not easily surpass simple univariate time series models.

### Interpreting machine learning results

This study presents the interpretation of the ML model's result, particularly the ML with the highest prediction performance, the XGB, using the SHAP value to explain feature importance. Our finding follows Lundberg et al. (2020), who advocate the gradient-boosted tree model for ML's result interpretation due to its high accuracy and low bias, both local and global, using the SHAP value.

#### *Feature importance*

Based on the XGB model’s result, as suggested by Lundberg et al. (2020), we apply the TreeExplainer to compute the Shapley value to determine the contribution of each explanatory variable on the model output for each data from the training dataset. We then measure the global contribution from each feature on the model output, which is ranked according to the mean value of all Shapley values across the sample data, which is called feature importance. Figure 6 shows the impacting features on the inflation model output using SHAP values.

---

**Insert Figure 6**

---

The rank of the feature indicates the contribution to the model prediction using global SHAP values derived from the XGB model. The y-axis displays all the examined features, while the x-axis displays the average magnitude of the SHAP values that illustrate the change in model output when we remove such a variable. The feature importance analysis reveals that ATM and debit cards (ATMD), credit cards (CC), and interest rates (IR) are the top three predictors that have significant contributions in predicting the inflation rate represented by the overall SHAP value in the XGB model.

The study discovered that several indicators typically employed for inflation forecasting or macroeconomic forecasting, such as the exchange rate (ER), the consumer confidence index (CCI), and the return of the capital market (PER), were found to be less significant than expected.

***Feature importance explanation***

Figure 7 shows the global SHAP values in a summary plot for the XGB prediction model based on inflation rate data. In the summary plot, a high value for individual data of a feature corresponds to a red dot, while a blue dot represents a low value. The x-axis displays the SHAP value of individual data in each feature's prediction across the observation periods, which measures the extent to which the removal of a feature affects the change in prediction. It should be noted that the interpretation of summary plots for classification and regression ML models may have some common ground but may also differ at some point.

In the case of a classification ML model that classifies the observed variable in a binary number, for instance, 0 does not have cancer, and 1 (one) is positive for cancer. The result can be clearly interpreted, for example, as an increase of feature X1 will increase the risk of a person being diagnosed with cancer. In contrast, the feature

X2 will have a negative effect on the probability that some individuals have cancer (Lundberg et al., 2019). By observing the concentration of red dots with the positive Shapley value of a feature in Figure 7, i.e. credit card (CC), we can conclude that a rise in the use of CC may lead to an increase in the inflation rate. In contrast, the cluster of blue dots in the positive Shapley value for the interest rate (IR) feature suggests that an increase in the IR may result in a decline in the inflation rate.

**Insert Figure 7**

However, is it reasonable to conclude that an increase in the usage of ATMD for transactions will lead to an increase in inflation? According to economic theory and most empirical studies, the increase in money or other forms of money resulted in a rise in the inflation rate (Warjiyo & Juhro, 2022). Consequently, time must be considered when interpreting the results of the ML regression model, as will be detailed in the following section. In addition, since the XGB method is a nonparametric ML model, unlike linear models, the relative importance of each feature in determining the observed variable might shift as the model learns from the data.

***Local Shapley Interpretation***

*1) Dependence Plot*

After we examine the importance of each feature and how each feature affects the aggregate model performance using global SHAP values, we then explore the level of importance of each explanatory variable across different observation points of the dataset using the local Shapley values. We use dependence plots to comprehend the relationship between the value of each explanatory variable and the model's predicted outcomes, which are represented by the SHAP values of each observed point, as shown in Figure 8.

**Insert Figure 8**

The outcome of the dependence plot depicted in Figure 8 can be interpreted as follows: SHAP values above the y-axis = 0 (positive) indicate a predicted increase in inflation, whereas SHAP values below the y-axis = 0 (negative) indicate a decrease in the prediction of inflation. As shown in Figure 8a, the dependence plot of interest rate demonstrates a negative slope, from positive to negative SHAP values, with a few outliers; therefore, an increase in interest rate led to a fall in the inflation rate. The point at which the distribution of SHAP values crosses the y-axis = 0 represents the threshold at which the model's prediction changes from an increase to a drop in the inflation rate, with the change occurring at an interest rate of approximately 6.6%. In contrast, Figure 8b shows that the credit card has a positive slope, and most of the dots have a positive value, indicating a positive relationship with the inflation rate; an increase in credit card transactions leads to an increase in inflation. This outcome is consistent with previous empirical research.

**Insert Figure 9**

Attempting to interpret the relationship between each predictor and the observed variable using a dependence plot can be challenging due to the limited amount of data included and the ambiguous pattern of the scatter plot. For instance, in the case of ATMD shown in Figure 9a, the SHAP values are predominantly negative throughout the periods, indicating a negative association with the inflation rate. The ATM or debit card provides access to a customer's savings account and the ability to make purchases easily; hence, it can be considered a type of money and consequently can positively affect inflation (Aprilianto, 2022).

However, Lubis et al. (2019) use debit cards (and other card payments) as a proxy for payment instrument innovation and found that the debit card replaces the cash function for transactions; hence, it negatively influences money demand. The dependent plot of currency exchange-sale transactions (KUPVAs), as represented in Figure 9b, reveals that the plot of SHAP values is randomly dispersed, presenting an ambiguous pattern. Considering the inconclusive result of the dependence plot of ATMD and KPUVAs, we construct a functional form of each feature's SHAP values and the observed value, following Buckmann et al. (2022), for an in-depth analysis.

*2) Feature Functional Form*

As indicated in the bee swarm plot in Figure 7, the functional form is built using local Shapley decompositions acquired by the XGB model. The functional form maps the local Shapley values of each feature individually to their observed values and visualises their relationship. Local Shapley decompositions were plotted using the best-fitting polynomials of the first and second degrees.

**Insert Figure 10**

For instance, Figure 10 demonstrates the functional structure of ATMD and KUPVAs in the inflation prediction model. The data indicate that ATMD has a positive slope, indicating that an increase in the use of ATM and Debit cards leads to a rise in the inflation rate. This result is consistent with economic theory and prior economic research. The functional form of ATMD and inflation rate provides a better explanation, supplementing the dependence plot as previously shown in Figure 9a. In contrast, the slope for KUPVAs is flat, as illustrated in Figure 10b, indicating that its influence on inflation is negligible. The results align with the global SHAP value outcome presented in Figure 7, which suggests that KUPVAs is the second least significant component inside the model.

**Insert Figure 11**

Figure 11 depicts other examples, including credit cards and interest rates. The functional form of credit cards (CC) suggests that rising credit card use is associated with rising inflation, which corroborates Wong and Tang (2020). Figure 11b depicts the interest rate (IR) functional form, which suggests that increases in the interest rate led to increases in inflation. This finding, however, contradicts the conventional literature on monetary policy and most empirical studies, which hold that an increase in the interest rate due to a contractionary central bank policy decreases inflation and economic activity in the short-run, with a lagged effect (Juhro et al., 2021, Maehle, 2020).

Because the result of the dependent plot and the functional form for interest rate were contradicts the economic theory and most previous research, further examination is needed, particularly in the technical aspect with the outliers. Following Buckmann et al. (2022), we then remove the extreme input values to resolve the issue. Figure 12 shows the functional form of the interest rate after extreme values have been removed. The evidence suggests that interest rate has a negative impact on the inflation rate.

Accordingly, we conclude that the functional form graph strengthens the dependence plot when explaining the outcome of ML prediction, especially when trying to comprehend the relationship between each feature and the target variable.

---

**Insert Figure 12**

---

*3) Interaction between features*

We extend the use of dependence plots to capture not only the relationship between the SHAP value of each feature and the model's predicted outcomes (inflation) but also the interaction between the observed variables. As an illustration, we compare the dependence plots of interest rates (IR) with and without interactions with the price-earning ratio (PER) and the currency demand (CIC) in the XGB model equation, as shown in Figure 13. Figure 13a depicts an interest rate dependence plot, with interest rate values shown on the x-axis and SHAP values (or model influence) indicated on the y-axis. Figures 13b and 13c illustrate the interaction of interest rates with currency demand (CIC) and the price-earnings ratio (PER), respectively, based on each feature's attribution to the model's forecast.

---

**Insert Figure 13**

---

We extend the use of dependence plots to capture not only the relationship between the SHAP value of each feature and the model's predicted outcomes (inflation) but also the interaction between the observed variables. As an illustration, we compare the dependence plots of interest rates (IR) with and without interactions with PER and CIC in the XGB model, as shown in Figure 13. Figure 13a depicts the IR dependence plot, with interest rate values shown on the x-axis and SHAP values (or model influence) indicated on the y-axis. Figures 13b and 13c illustrate the interaction of IR with CIC and PER, respectively, based on each feature's attribution to the model's forecast. In the interaction dependence plot, each data point represents individual data in a distinct period of observations, coloured differently based on connected parameters, which are CIC and PER, at the high and low values. Figure 13b reveals that most data points with a notably low-interest rate correspond to higher currency demand (red dot); these data points have positive SHAP values. As the interest rate increases, SHAP values of IR decrease, and this relationship continues linearly as more data points with negative SHAP values. We discovered that the density of CICs with low values (blue dot) increases when the interest rate is high, despite the occurrence of a few CICs with high values. This finding corroborates the quantity theory of money and previous studies (Lim & Dash, 2021; Lubis et al., 2019).

In the case of the PER, most data points with low interest rates correspond to a higher stock market earning ratio (red dot) with positive SHAP values, and vice versa. PER tends to decrease as the interest rate increases and SHAP values reduce. As demonstrated in Figure 13c, the negative relationship between IR and PER continues linearly to a more data point of low PERs (blue dot) gathered as SHAP values become negative, which corroborates Zhao and Bacao (2021).

**Further discussion**

Forecast error minimisation is the primary objective of ML models; this study discovered that the ML models successfully reduced the forecast error on average by 45.16% compared to ARIMA as the benchmark, with the XGB model performing the best. However, the performance of both ensemble learnings decreases when used to make longer-term predictions (24 months). Therefore, while employing a regression ML model for prediction, it is crucial to consider the prediction span and data splitting ratios.

Using global SHAP values, this study exhibits that seven of the top ten most influential predictors are payment system variables; therefore, it is reasonable to conclude that payment system indicators are critical measures for inflation forecasting. Our findings are partially consistent with those of earlier studies that used data from payment systems for macroeconomic forecasting. For instance, a study by Aprigliano et al. (2019) demonstrated that payment system transaction data can track economic activity and improve macroeconomic forecasting. It was also confirmed by Chapman and Desai (2020) for forecasting macroeconomic indicators using payment system data during the pandemic period. However, its performance may vary slightly depending on the nowcasting scenario. We perform further examination using local SHAP value decomposition and dependence plot and explain how each explanatory variable affects the inflation rate, as indicated by its contribution to the model prediction in the XGB model. We then use the functional form plot to emphasise the dependence plot's result and highlight the relationship between each predictor and inflation.

Based on the dependence plot and functional form plot, we found that ATMD and credit card transactions positively influence the inflation rate. In contrast, interest rates negatively affect inflation. These results align with those of earlier empirical research mentioned above. To give a more accurate interpretation, we recommend removing some outlier data, as shown in Figure 12, in examining the impact of interest rates on inflation and increasing the number of data input into the model. As seen in Figure 7, the Global SHAP value reveals that some variables are insignificant to the model. For example, the functional form graphs of Figure 10b and Appendix C (Figure i) provide a flat slope illustration of KUPVAs and the delivery channel (DC). Another example is the exchange rate (ER); we found that an increase in the exchange rate leads to an increase in the inflation rate (Appendix C, Figure g), which corroborates with previous economic studies (Isnowati et al., 2020, Amhimmid et al., 2021). The examination of the result of the ML prediction can also be extended using an interaction dependence plot, such as how an interest rate interacts with other predictors based on their attribution to the model prediction.

In addition to outliers' exclusion, the statistical independence assumption of explanatory variables is another technical aspect of ML interpretation that should be considered. Violating this assumption may lead to biased coefficient values, erroneous predictions, and distorted Shapley values, as Aas et al. (2021) suggested. The above discussion of how to interpret ML results sheds light on ML's data-driven nature and ability to comprehend complex structures in economy-related issues. As a result, we believe that exploring ML models in the economic field would broaden the tools accessible to academics and policymakers for handling more complicated issues and comprehending more complex data in today's digital world.

We favour using ML as an alternative method for economic forecasting. However, we believe that in order for the ML used to be technically consistent and appropriately interpreted, especially for applied economics and policy implication analysis, the results must be made interpretable by adhering to guiding principles in forecasting, such as the classical assumption in econometric regression and the robustness check and discussed its results with previous empirical studies.

# CONCLUSION

This paper examines several prominent machine learning (ML) algorithms in estimating Indonesia's inflation rate, utilizing various payment systems, capital markets, and macroeconomic data. Overall, our findings indicate that all ML models produce forecasts with lower RMSEs than the ARIMA benchmark and can reduce average forecast errors by 45.16 % compared to the benchmark. The RMSE, MAE, and Diebold-Mariano tests demonstrate that the XGB has the best forecasting performance. Generally, the accuracy of each algorithm varies with the size of the data split, and the forecast error of the model's predictions decreases as the ratio of testing data to training data decreases; however, RF and XGB regression experience a worsening forecast error in the more extended period of data (24 months).

The feature importance analysis employing global and local SHAP values yielded several interesting results. First, ML forecast results can be appropriately interpreted using several tools, such as feature importance, dependence plots, and functional forms, which complement one another to provide a more accurate causal inference. The expansion of local Shapley decomposition could also shed light on the relationship between the explanatory variables within the ML model using the interaction dependence plot. However, caution is required in the interpretation, which must adhere to econometric forecasting principles and be compared to existing economic theory and empirical studies. Second, seven of the ten most influential predictors are payment system variables, indicating that payment system indicators are essential for inflation forecasting.

Finally, this study contributes to the growth of interdisciplinary research in applied economics and computer science, which can deliver more vigorous knowledge to the practical field than discipline-specific research alone. Our findings advocate the adoption of ML algorithms as complementary tools to assist the central bank in anticipating future conditions and comprehending its primary objective of maintaining price stability. This study encourages and supports further research in applying more ML models to determine the causal relationship between economic variables, including incorporating more contemporary features, both structured and unstructured data, with longer and more granular data.

## ACKNOWLEDGEMENT

**Acknowledgements:** The author would like to thank all those who helped me during this research project. The author would like to express gratitude to his supervisors, Prof. John Fender, Prof. Yiannis Karavias, and Prof. Christoph Gortz, the Head of Payment System Policy Department of Bank Indonesia, the Head of Bank Indonesia Institute, and DR. Asrul Harun Ismail, for introducing me to machine learning and sharing your insights on its application in industry and business.
**Funding:** This work was supported by the LPDP Indonesia.
**Conflicts of Interest:** The authors declare no conflict of interest.

## Tables and Figures

Figure 1. Time series of some selected variables

a) Inflation rate

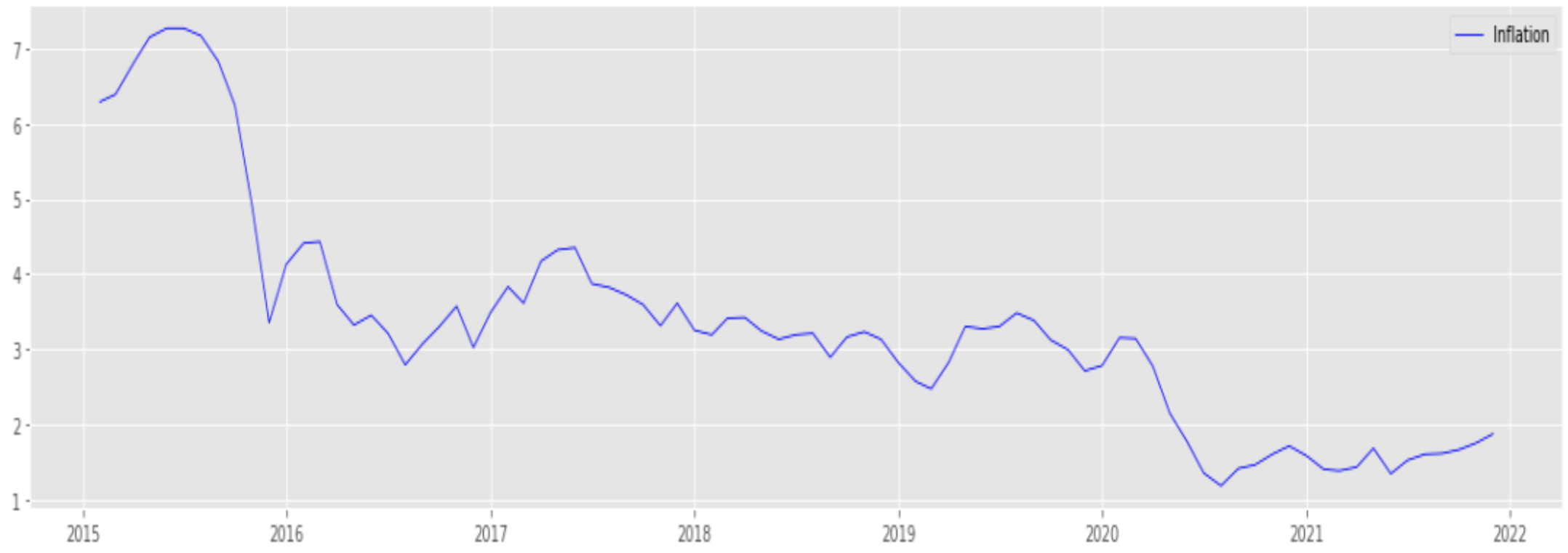

b) Change of cash in circulation, ATM/Debit card, average daily trading, and fund transfer from abroad (all in log)

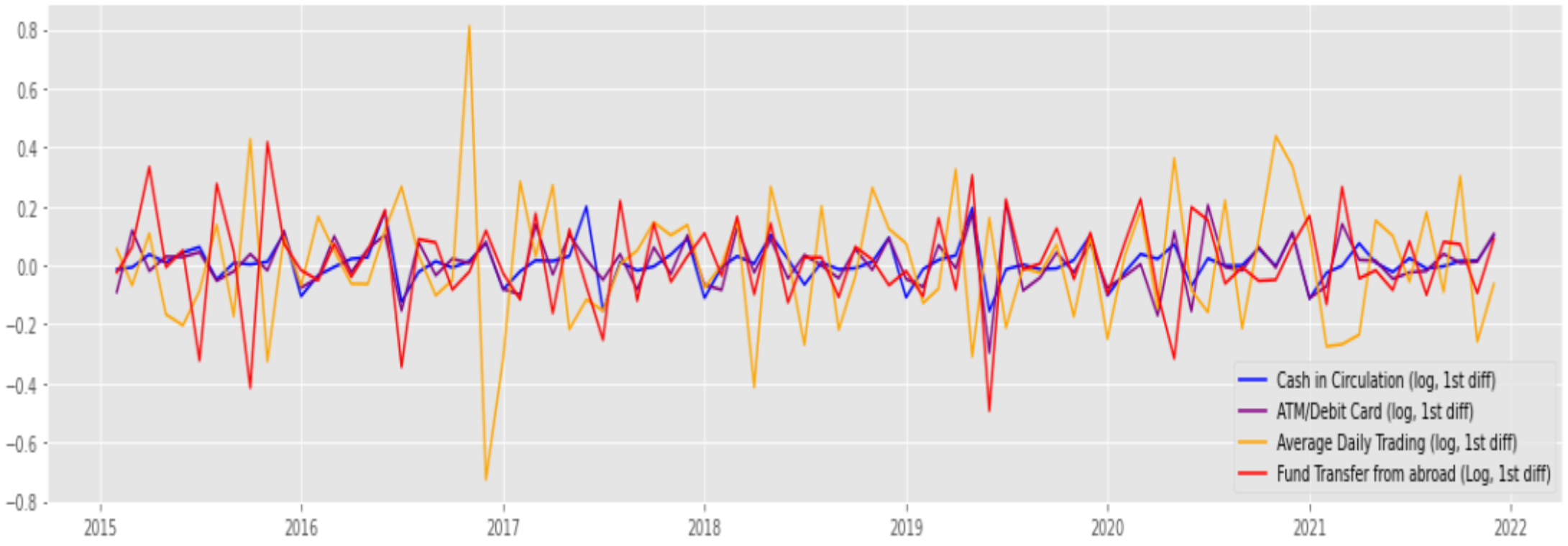

c) Credit card transaction, currency exchange-sell, and price earning ratio (all in log)

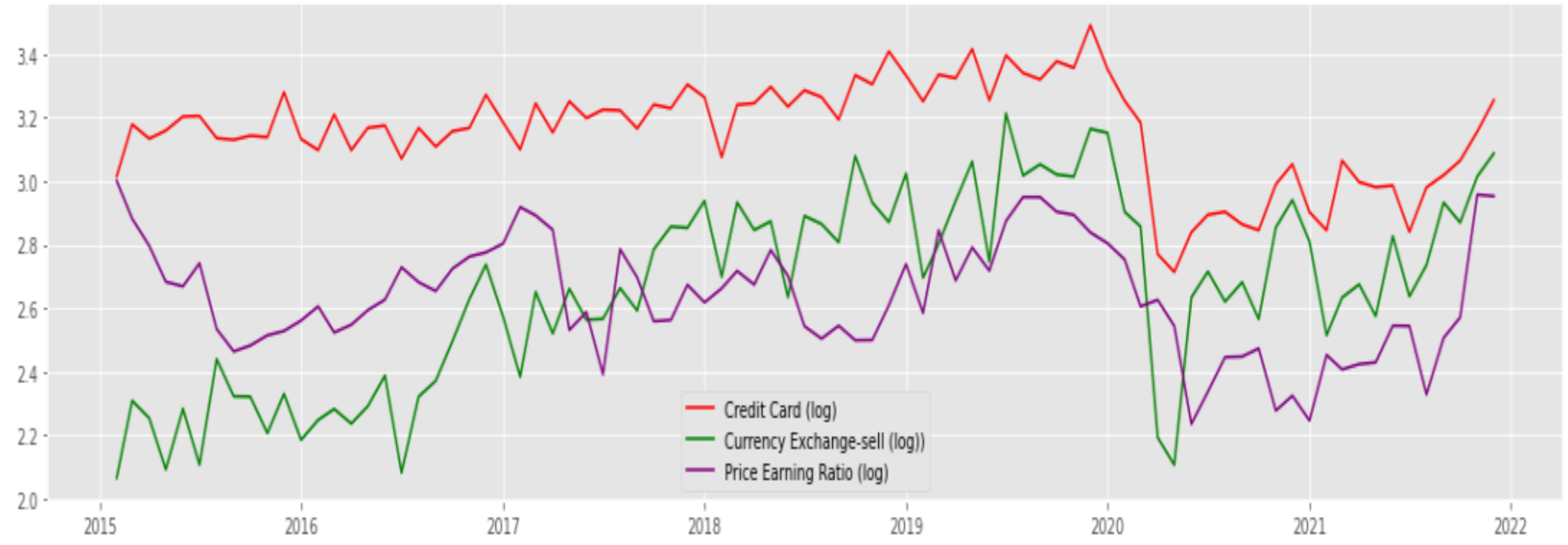

*Source: Bank Indonesia*

Figure 2. The procedure of K-fold cross-validation in the training data set and finding model parameters.

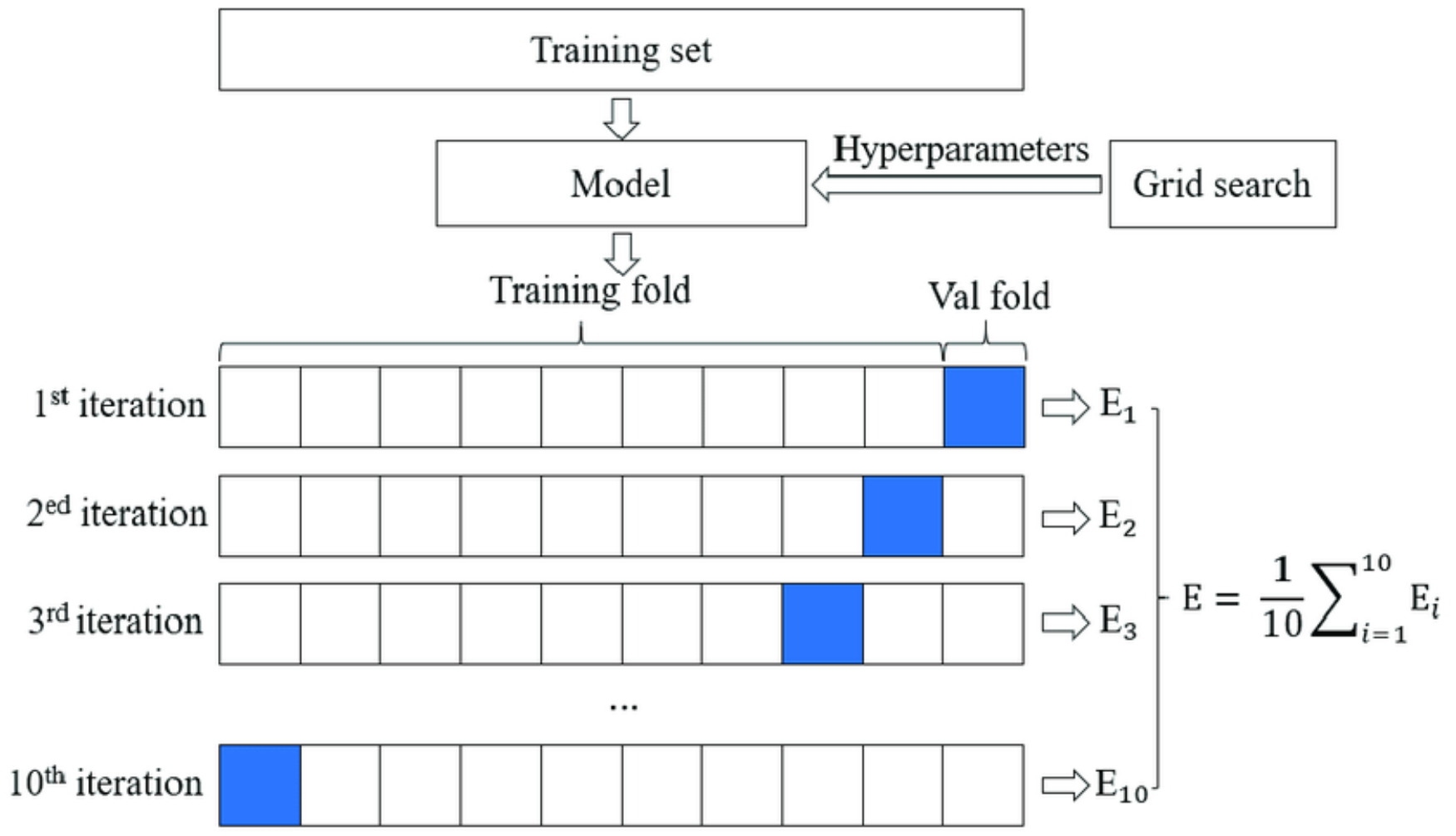

Source: Guo et al. (2021)

Figure 3. Out-of-sample prediction of the inflation rate of Indonesia of all ML models, 2020m9 to 2021m12

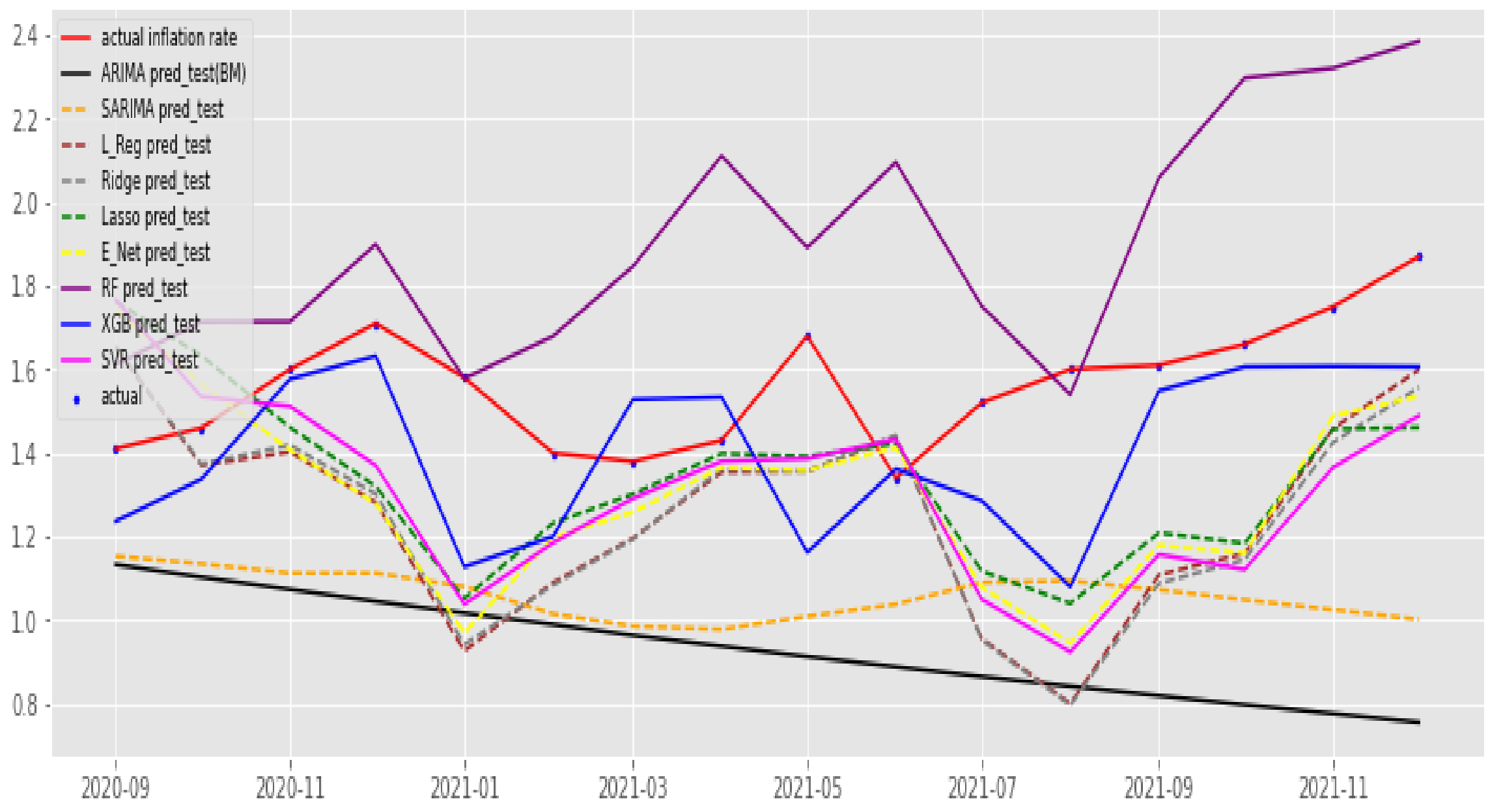

*Source: Author's calculation*

Figure 4. The inflation rate of Indonesia (log) for the period from 2015:M1 to 2021:M6 and forecast of XGB regression with forecast error

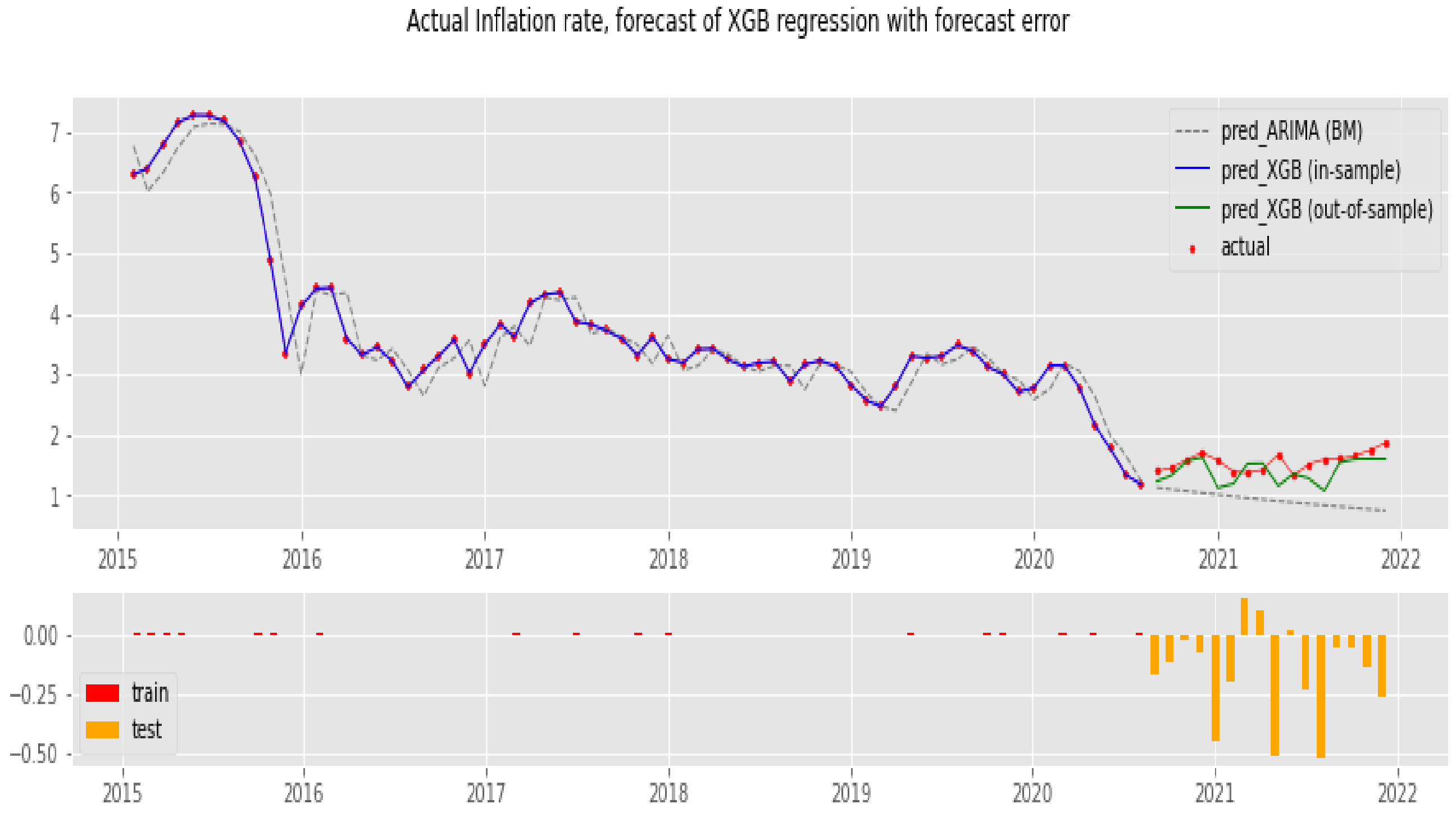

Figure 5. RMSE of the examined models for various out-of-sample prediction periods (24, 16, 12, 9, and 6 months)

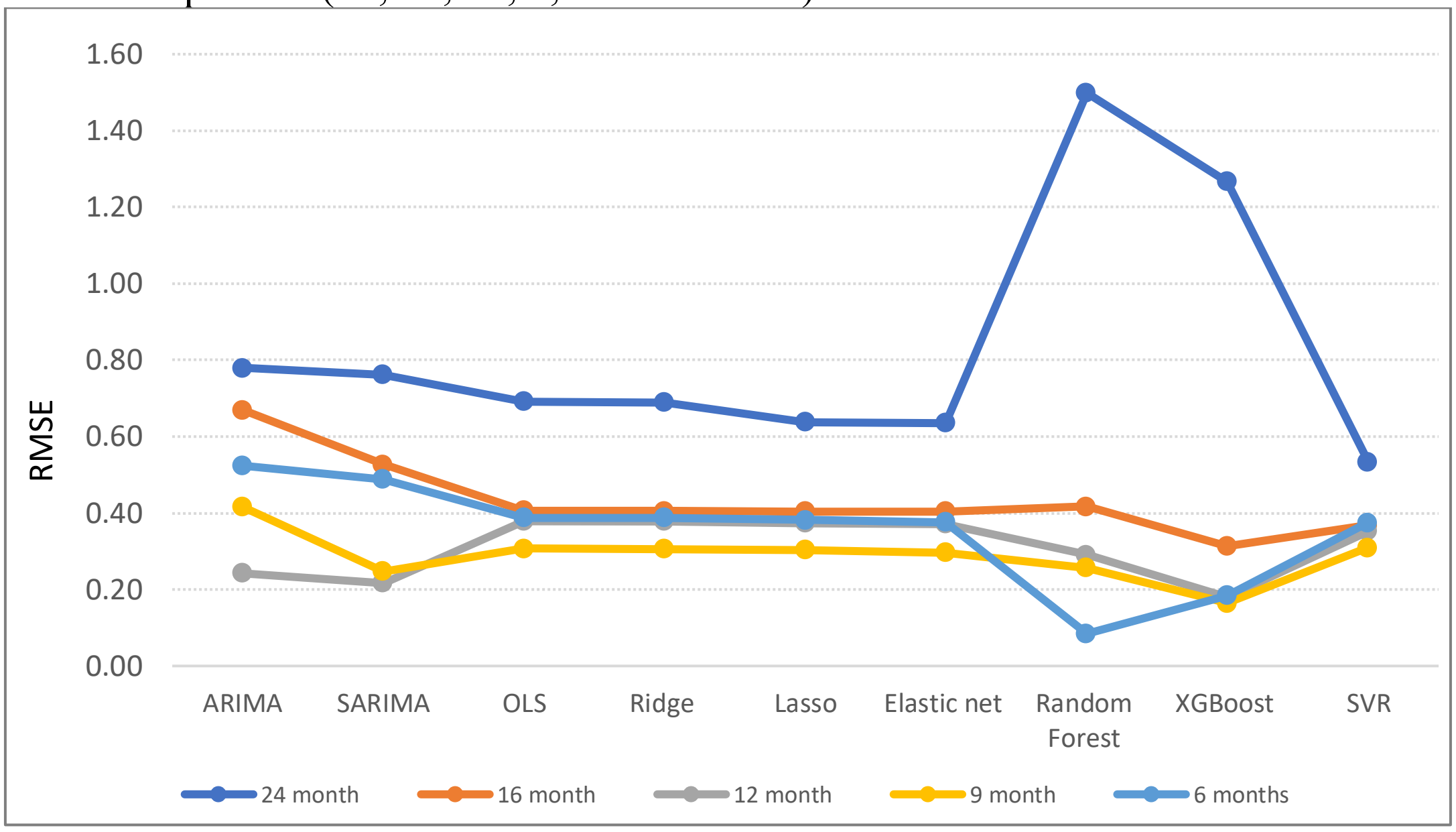

Figure 6. Feature importance of inflation using SHAP value based on XGB prediction. Source: Author computation

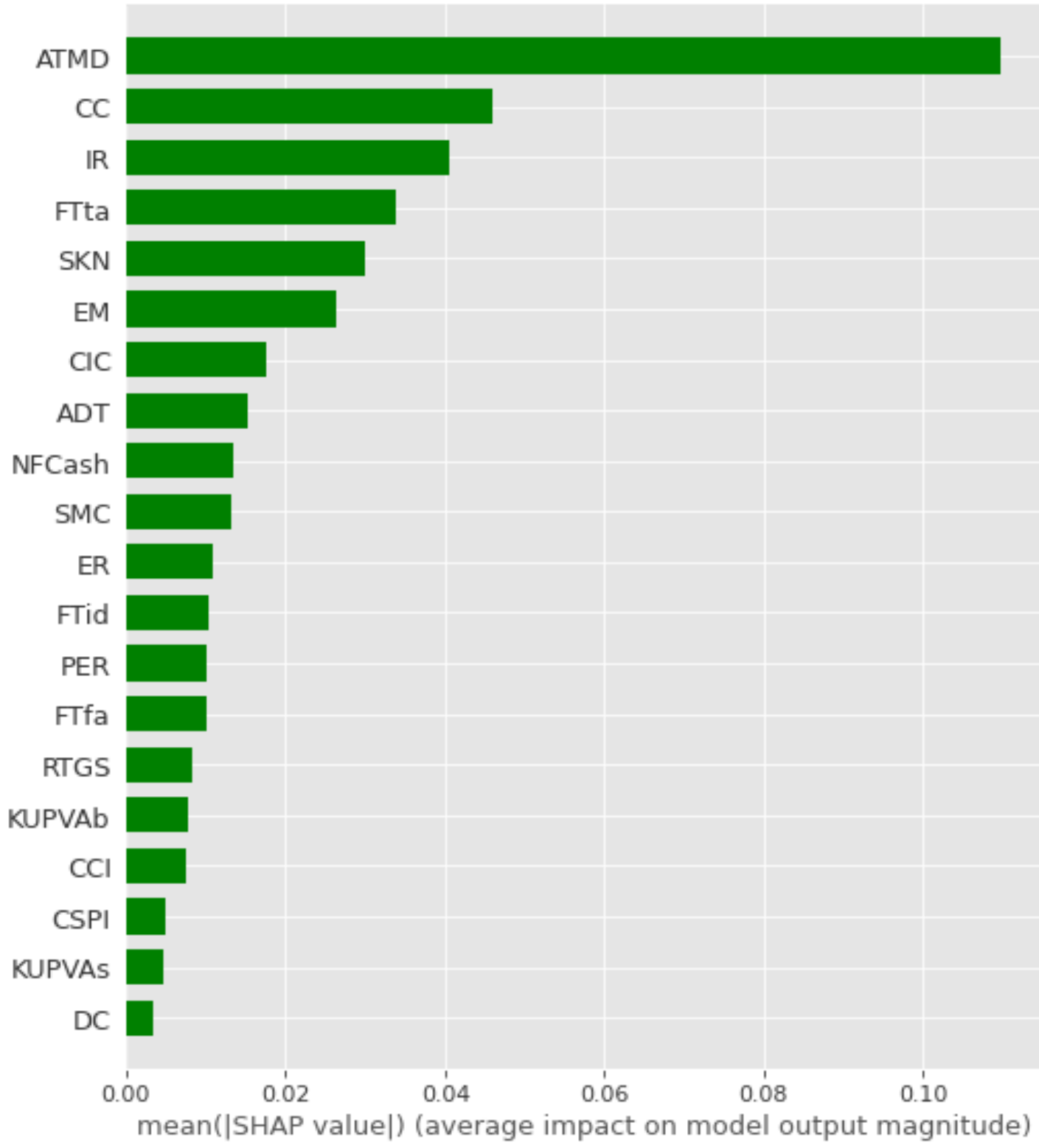

Figure 7. Summary plot of SHAP value of Inflation dataset based on XGB Prediction. Source: Author computation

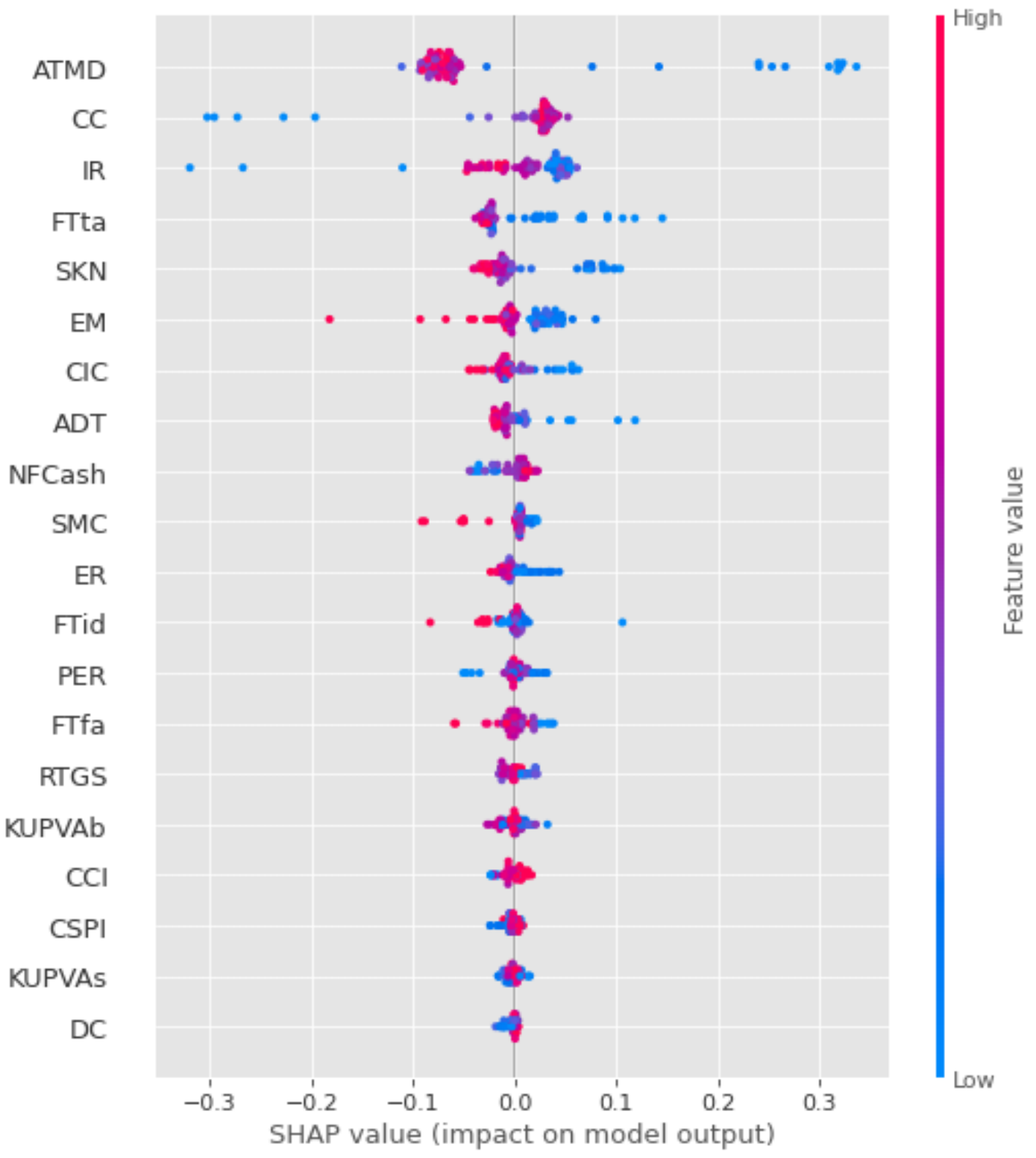

Figure 8. Dependence plot of SHAP value of Interest rate (IR) and credit card usage (CC). Source: Author's computation

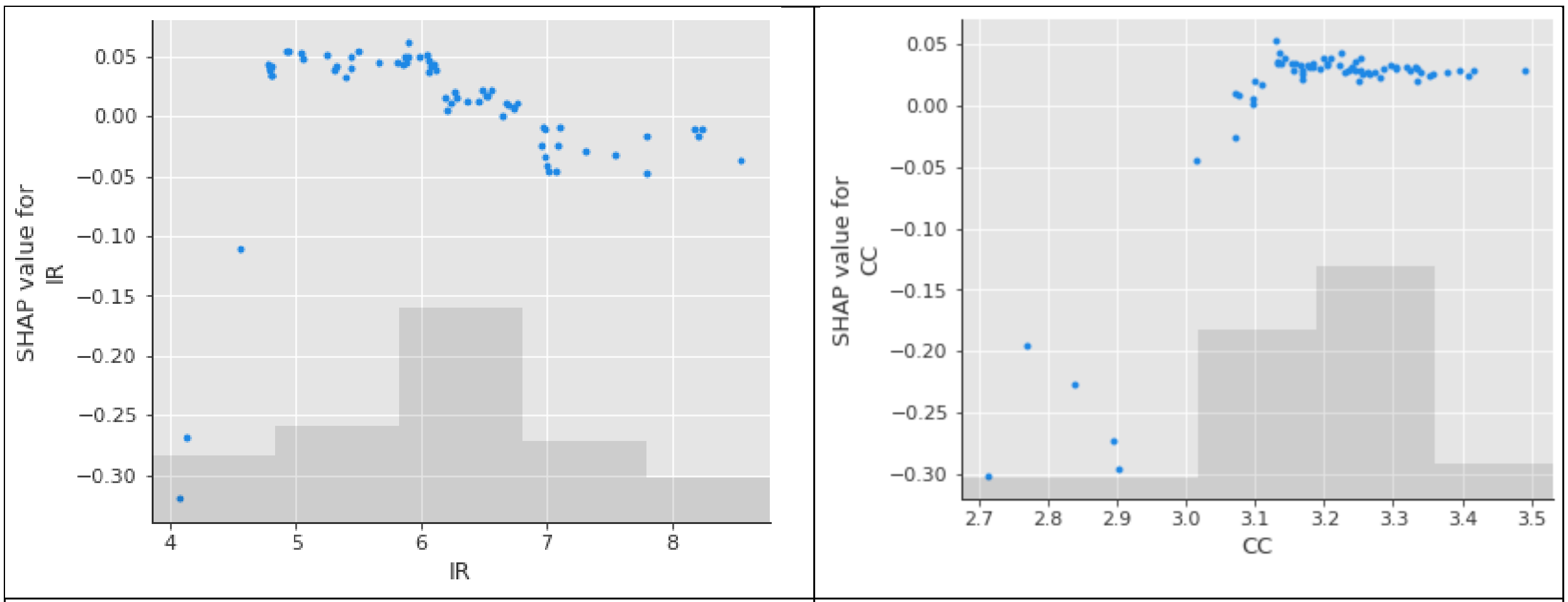

a) Dependece Plot of Interest Rate (IR)

b) Dependece Plot of Credit Card (CC)

Figure 9. Dependence plot of SHAP value of ATM and Debit Card (ATMD) and currency exchange-sale transactions (KUPVAs). Source: Author's computation

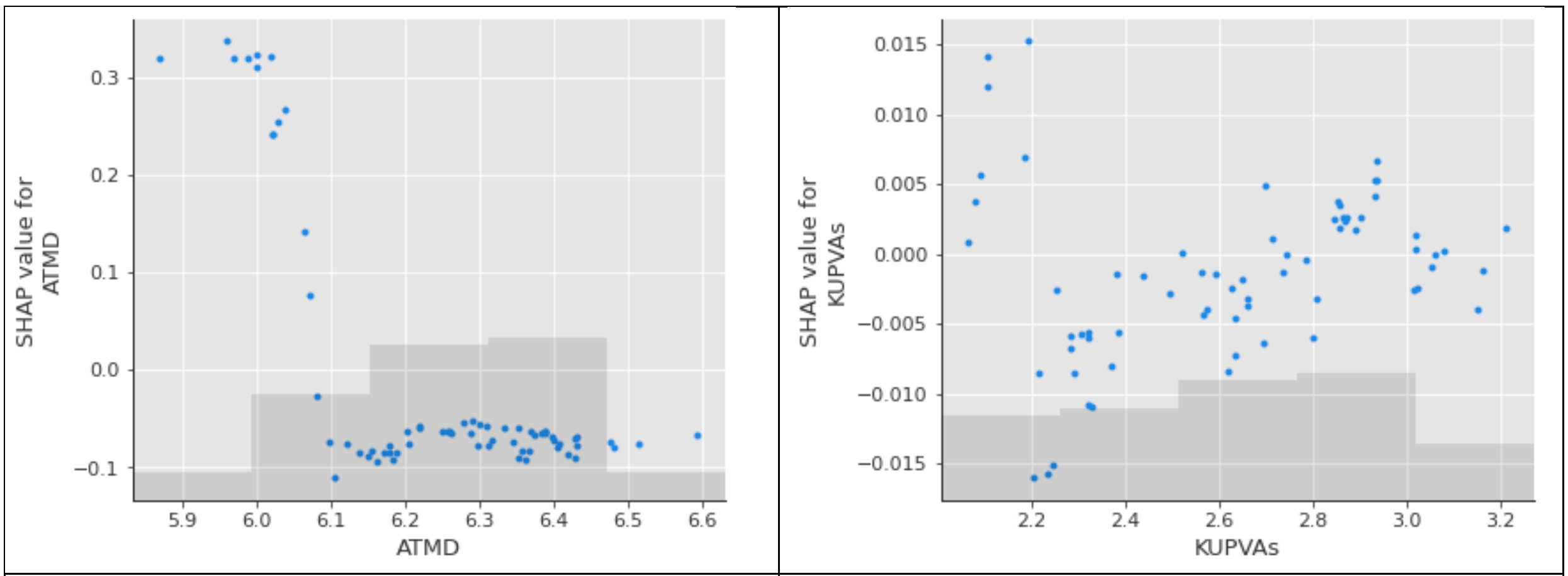

a) Dependece Plot of ATM and Debit Card (ATMI

b) Dependece Plot of currency exchange-sale transactions (KUPVAs)

Figure 10. Functional Form of SHAP value of ATM and Debit Card (ATMD) and currency exchange-sale transactions (KUPVAs). Source: Author's computation

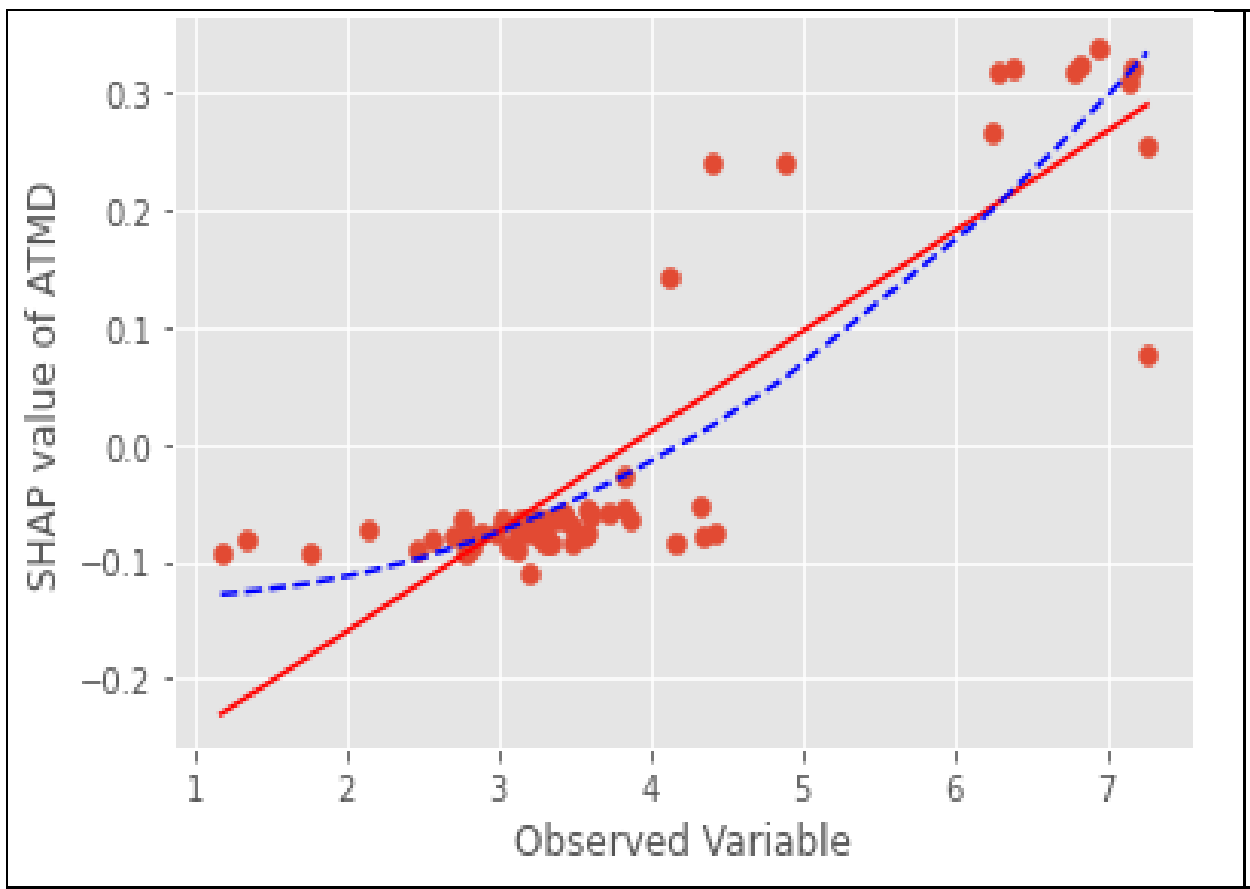

a) Functional Form of ATM and Debit Card (ATM

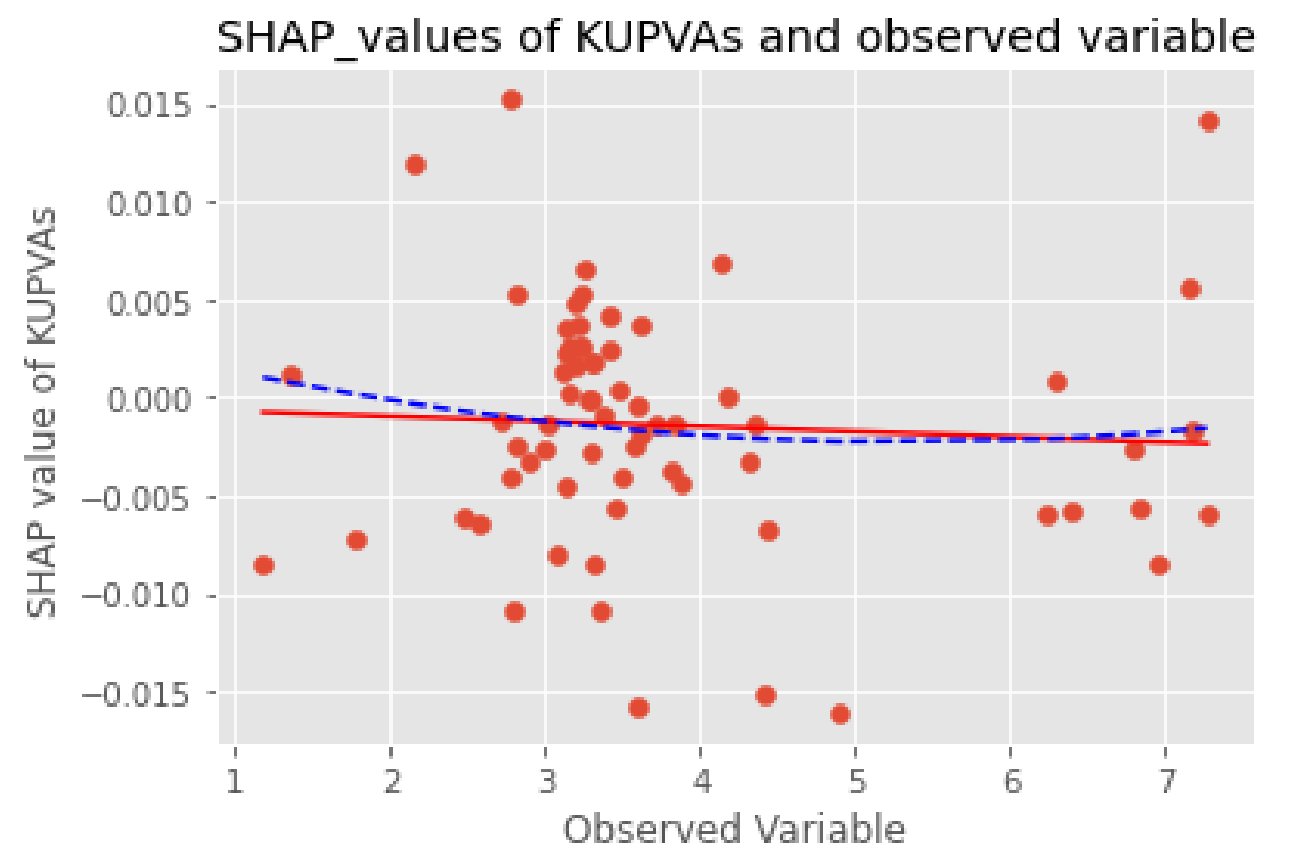

b) Functional Form of currency exchange-sale transactions (KUPVAs)

Figure 11. Functional Form of Credit Card (CC) and Interest Rate (IR). Source: Author's computation

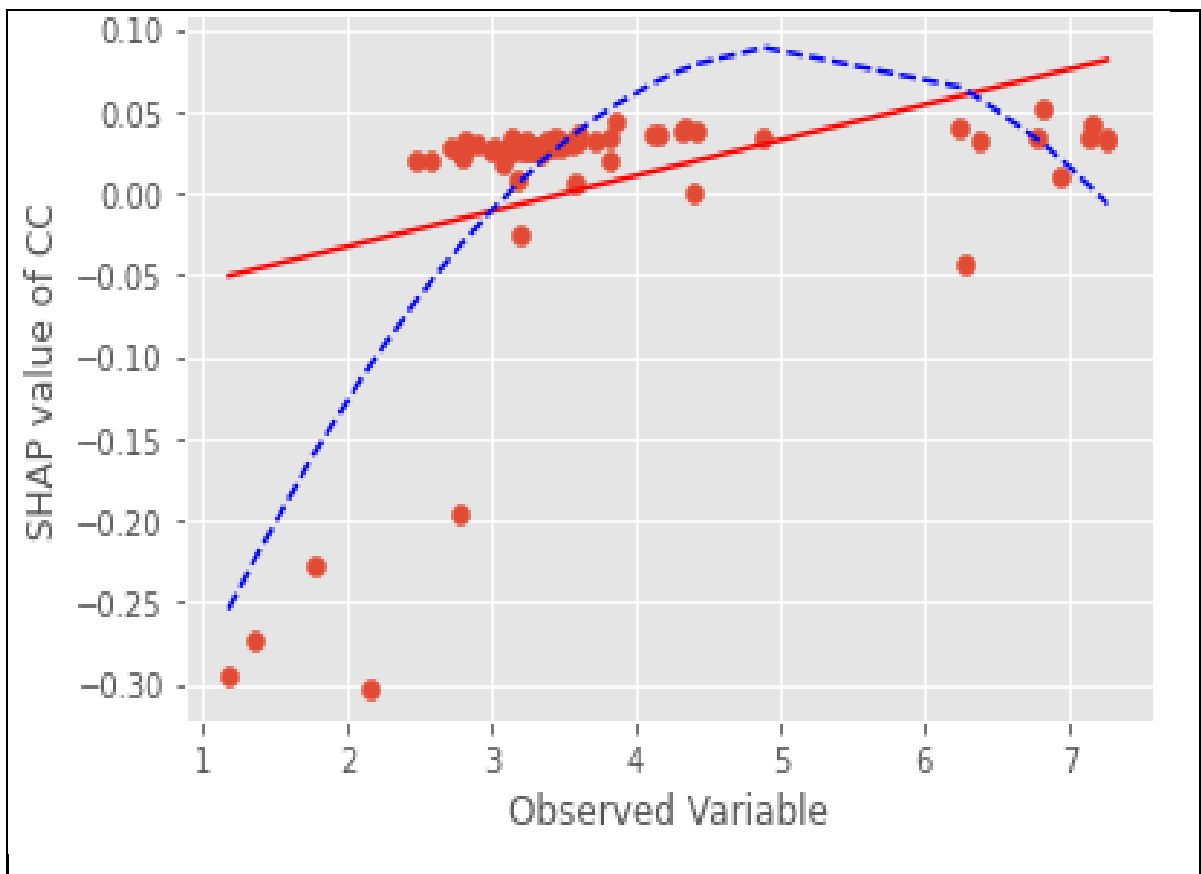

a) Functional Form of Credit Card (CC)

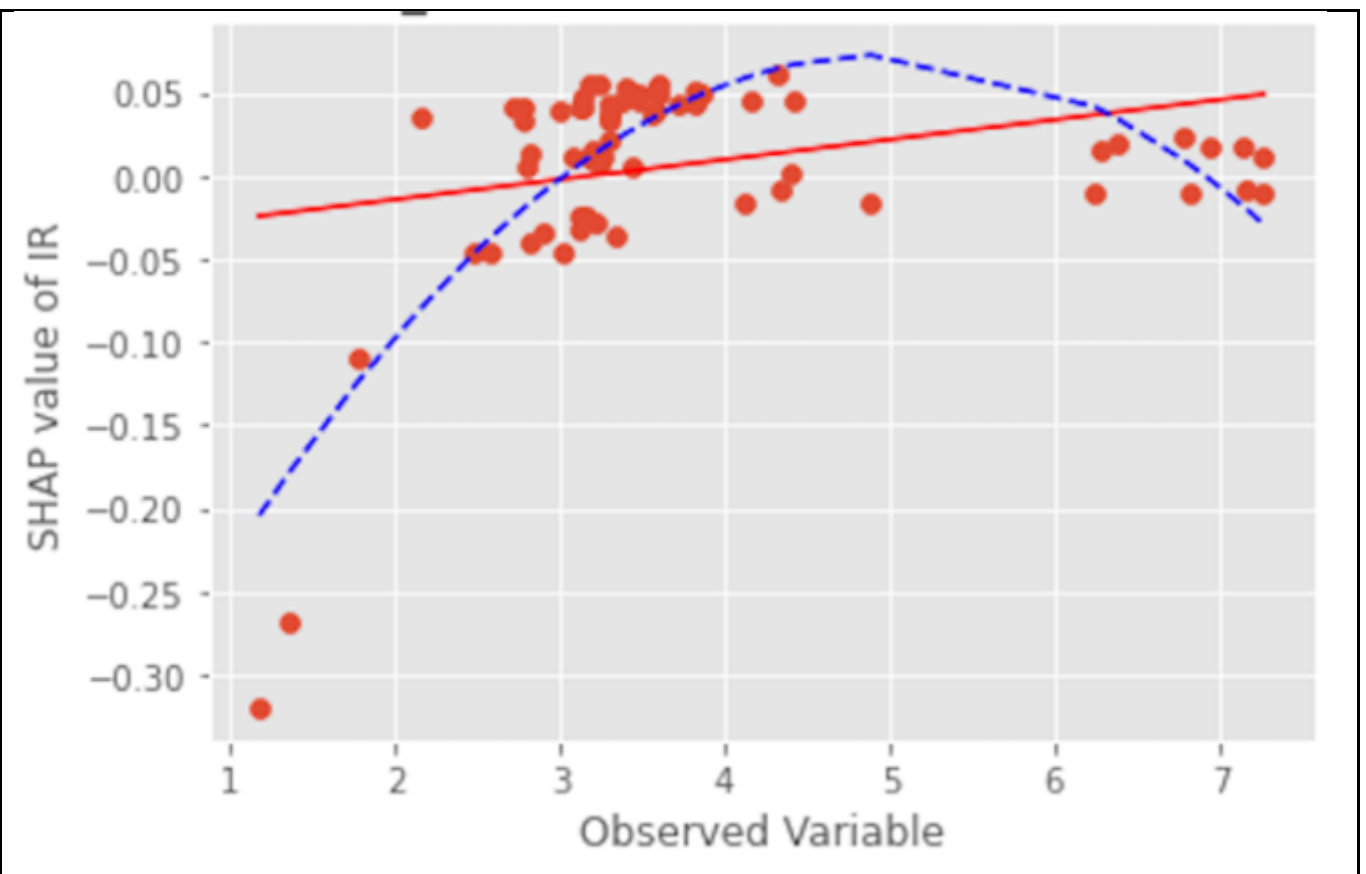

b) Functional Form of Interest Rate (IR)

Figure 12. Functional Form of Rate (after the removal of outliers). Source: Author's computation

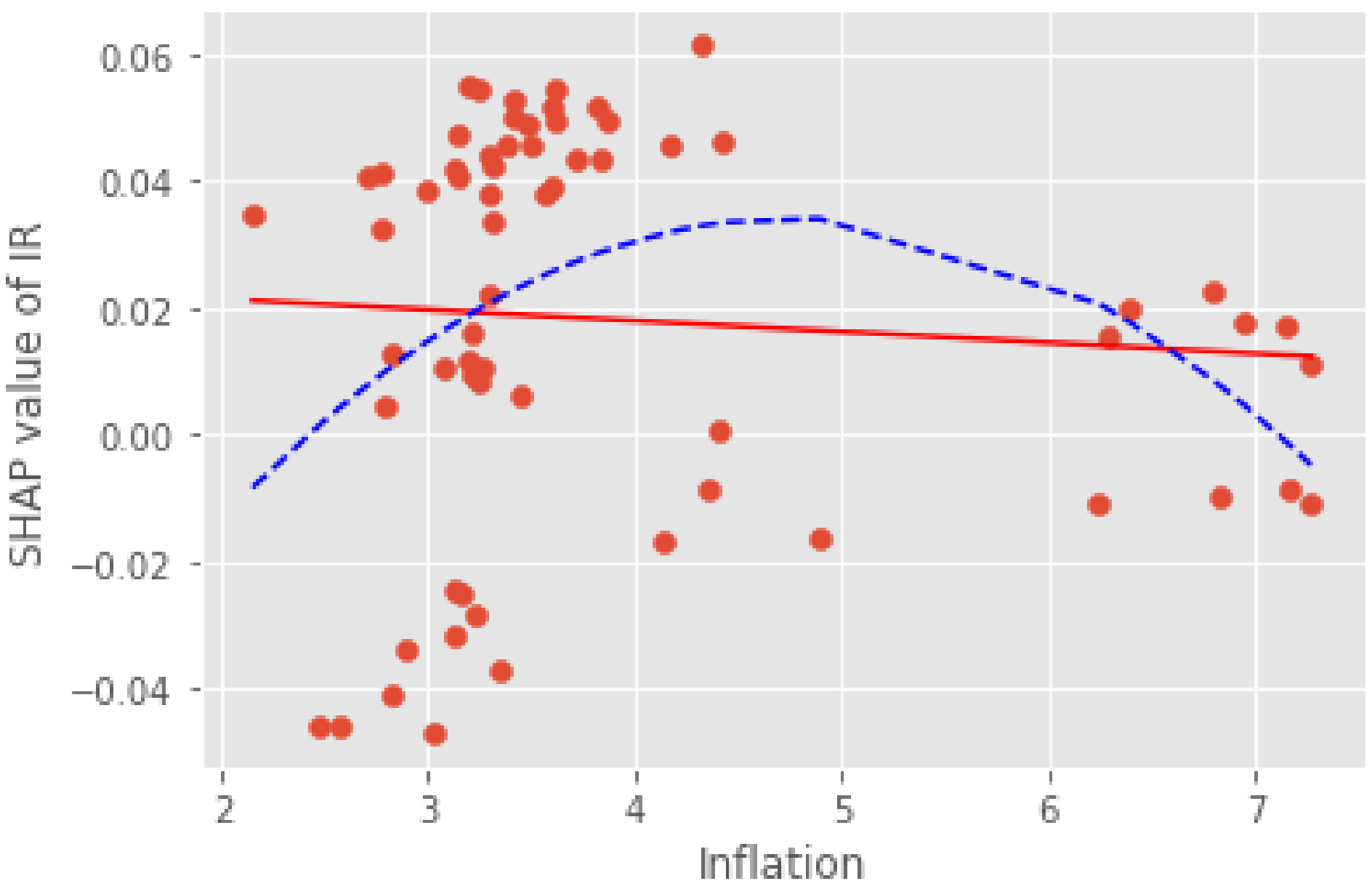

Figure 13. Dependence plot of Interest rate and interaction plot with selected variables (CIC and PER). Source: Author's computation

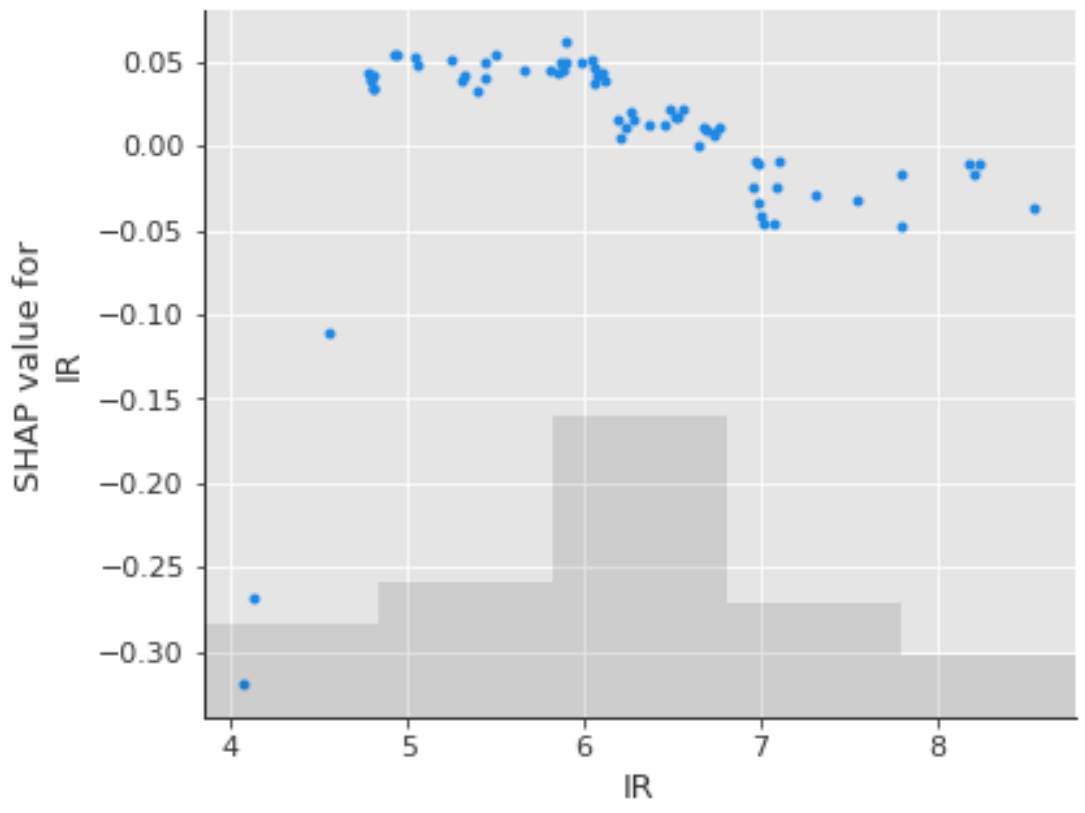

a) Dependence plot of SHAP value of Interest Rate (IR)

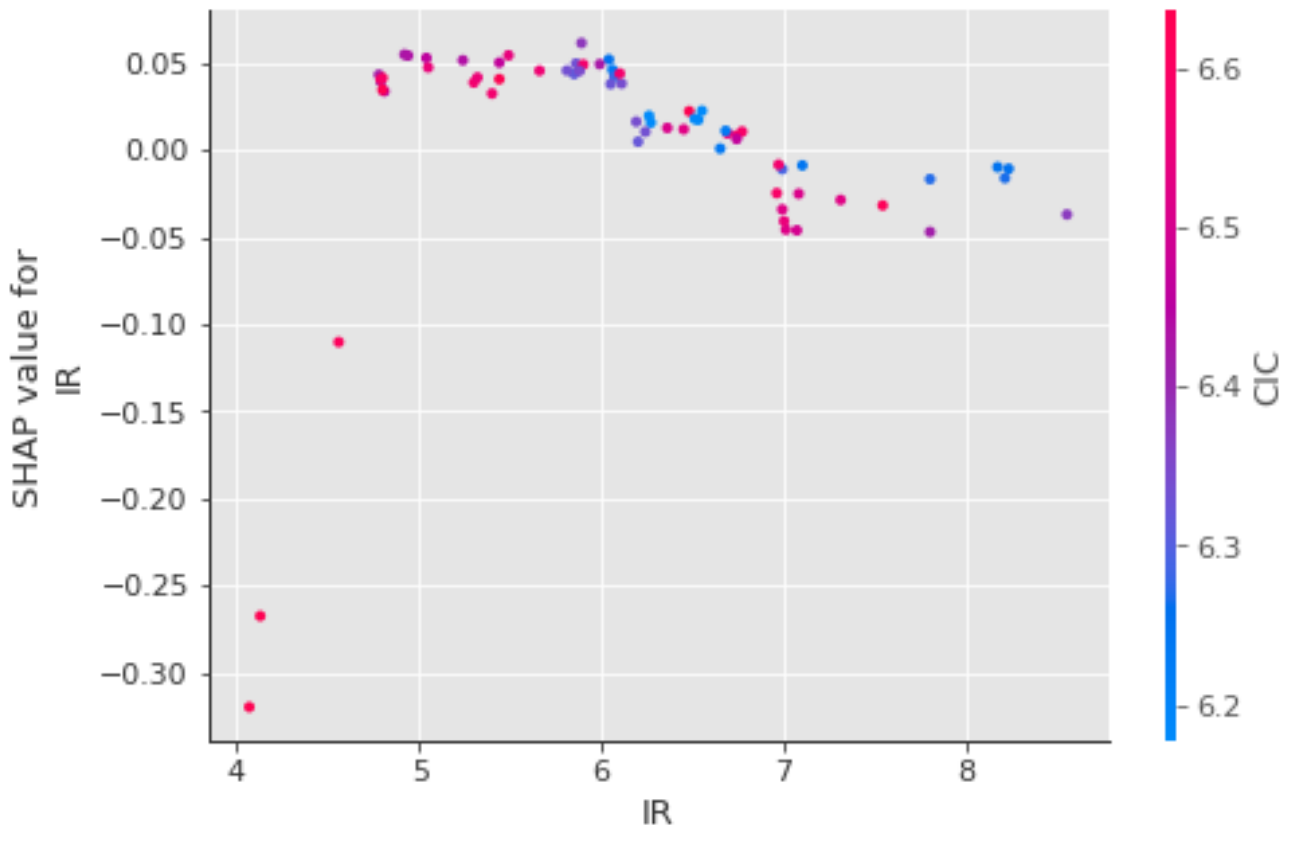

b) Interaction plot of IR and Cash in Circulation (CIC)

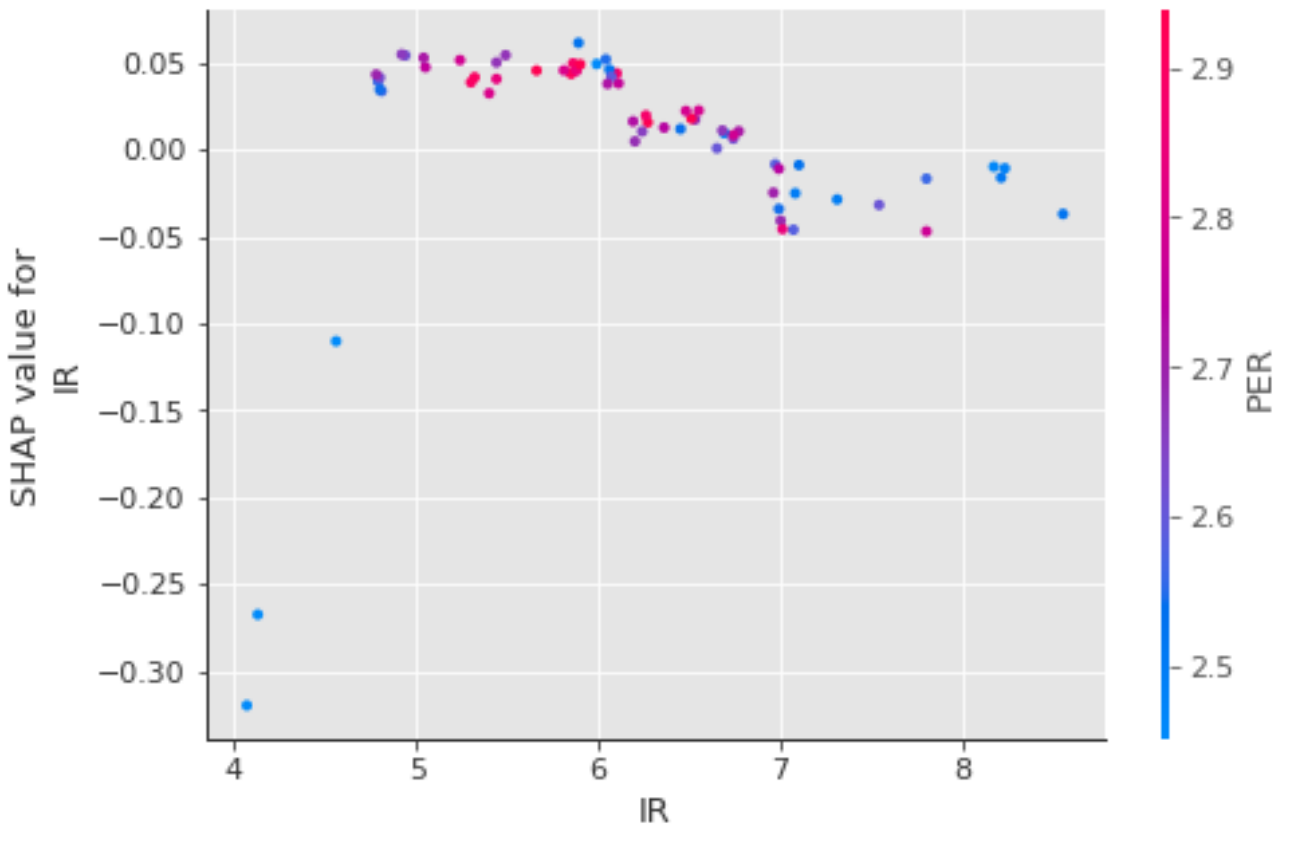

c) Interaction plot of IR and Price Earning Ration of Capital Market (PER)

Table 1. Hyperparameter Range and Value

| Model | Hyperparameter | Range | Optimised value |
|---|---|---|---|
| Ridge | Lambda | 0.001 to 0.9 | 0.001 |
| Lasso | Lambda | 0.001 to 0.9 | 0.001 |
| Elastic net | Lambda | 0.001 to 0.9 | 0.001 |
| | Alpha | 0.05 to 0.95 | 0.05 |
| Random Forest | Max Depth | 2 to 50 | 9 |
| | Max Features | 2 to 20 | 4 |
| | n_estimator | 10 to 1000 | 131 |
| XGB | learning_rate | 0.005 to 0.5 | 0.08 |
| | n_estimators | 10 to 1000 | 500 |
| | max_depth | 2, 4, 6, 8, 10 | 10 |
| | Subsample | 0.1 to 0.9 | 0.4 |
| | Colsample_bytree | 0.1 - 0.9 | 0.7 |
| SVM | C | 0.1 to 50 | 50 |
| | Epsilon | 0.0005 to 1 | 0.065 |
| | Kernel function | Linear/Polynomial/RBF | linear |

Tabel 2. Performance Metrics Comparison of ML Models and DM Test Related to the Comparison with Benchmark ARIMA Model, 2020m9 – 2021m12

| | **Performance Metrics** | | | **Statistic Test** | |
|---|---|---|---|---|---|
| **Model** | **MAE** | **RMSE** | **RMSE Reduction (%)** | **DM test (Rel. to ARIMA)** | **p-value (Rel. to ARIMA)** |
| 1. ARIMA (BM) | 0.630 | 0.670 | | | |
| 2. SARIMA | 0.503 | 0.527 | 21.31 | 3.988 | 0.001 |
| 3. OLS | 0.350 | 0.406 | 39.39 | 3.508 | 0.003 |
| 4. Ridge | 0.346 | 0.402 | 40.01 | 3.472 | 0.003 |
| 5. Lasso | 0.298 | 0.340 | 49.30 | 4.695 | 0.000 |
| 6. Elastic net | 0.317 | 0.364 | 45.62 | 4.216 | 0.001 |
| 7. Random Forest | 0.351 | 0.417 | 37.76 | 4.462 | 0.000 |
| 8. XGB | 0.264 | 0.313 | 53.22 | 4.875 | 0.000 |
| 9. SVR | 0.315 | 0.368 | 45.07 | 4.442 | 0.000 |

**Appendix A:** Data descriptions

| No. | Label | Short Description | Unit |
|---|---|---|---|
| **1** | RTGS | BI-RTGS transaction value comprises customer transactions, interbank money market, government, and monetary management transactions, among others. | Billions of Rp |
| **2** | SKNBI | National Clearing turnover is broken down by transaction type. It includes credit clearing (bulk and individual credit transfers), debit clearing (bulk and individual checks, bilyet giro, and others), and pre-fund and pre-fund usage (for credit and debit clearing). | Billions of Rp |
| **3** | ATMD | ATM and debit card transactions include cash withdrawals, purchases, and intra- and inter-bank transfers. | Billions of Rp |
| **4** | CC | Credit card transactions include withdrawals of cash and purchases. | Billions of Rp |
| **5** | EM | Electronic money transactions are classified into the following categories: purchasing, initialization and top-up, transfer between e-money, cash withdrawal, and redeem. | Billions of Rp |
| **6** | DC | The delivery channel refers to transactions that are completed over the phone, via SMS, mobile banking, or over the internet. | Billions of Rp |
| **7** | FT | Money transfer by nonbank includes: i) international outgoing fund transfers (FTta), ii) incoming fund transfers or transfers from abroad (FTfa), and iii) fund transfers within the Indonesian territory (FTid). | Billions of Rp |
| **8** | KUPVA | The value of currency exchange transactions, comprises of: i) buy transactions (KUPVAb) and ii) sell transactions (KUPVAs). | Billions of Rp |
| **9** | CIC | Cash data include currency in circulation (Milačić et al.) and the difference between deposit and withdrawal transactions to the central bank (NFCash). | Billions of Rp |
| **10** | ER | Exchange rate (ER) of USD against Indonesian Rupiah | Rp |
| **11** | IR | One month money market interest rate (IR), taken from Jakarta Interbank Offered Rate (JIBOR) | % |
| **12** | CSPI | Composite Stock Price Indexs, end period | point |
| **13** | SMC | Stock Market Capitalization | Billions of Rp |
| **14** | ADT | Average Daily Trading | Billions of Rp |
| **15** | PER | Price Earning Ratio (PER), average | times |
| **16** | CCI | Consumer Confidence Index | point |

**Appendix B:** Dependence plot of all explanatory variables, Source: Author's calculation

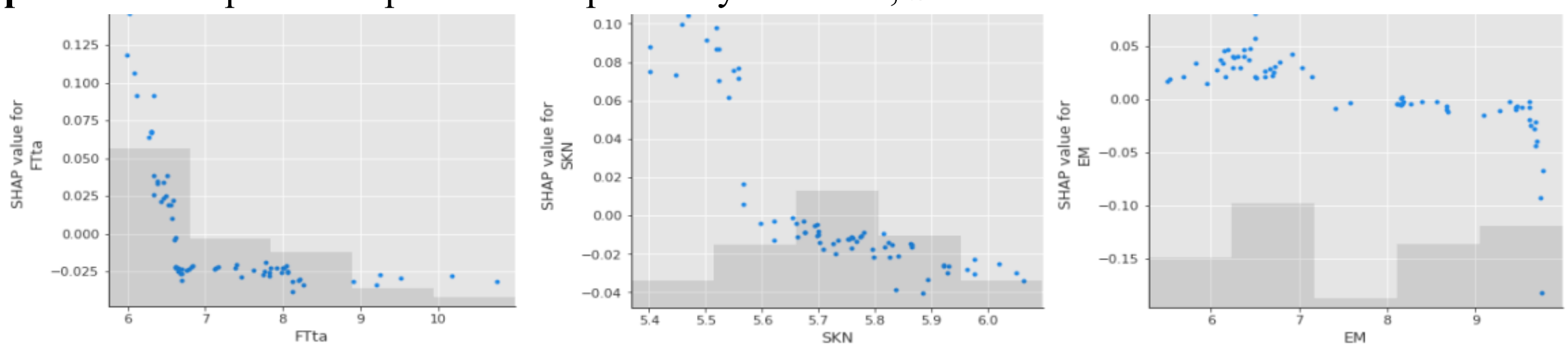

a) Fund Transfer to abroad (FTta) b) National Clearing (SKN) ) Electronic Money (EM)

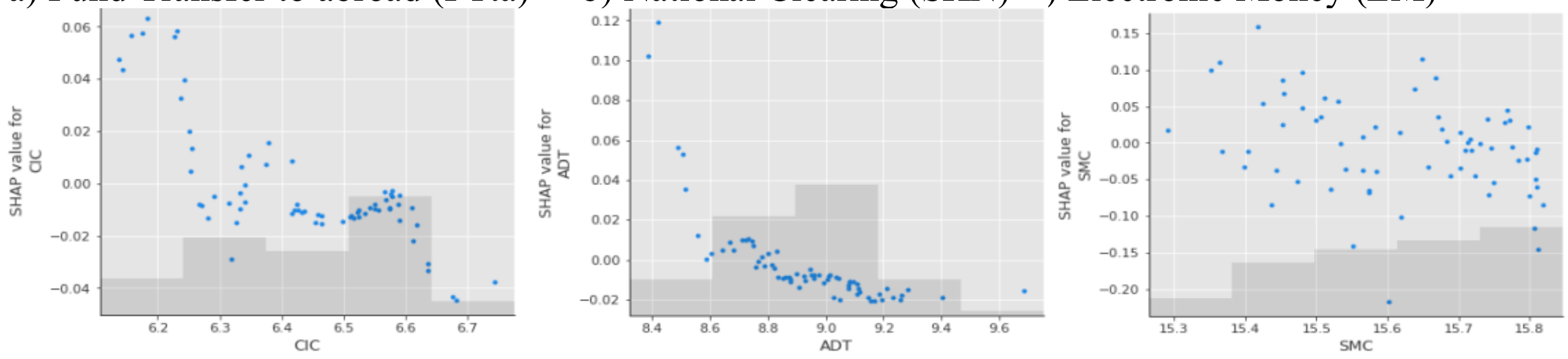

d) Cash in Circulation (CIC) e) Average Daily Trading (ADT) f) Stock Market Caps. (SMC

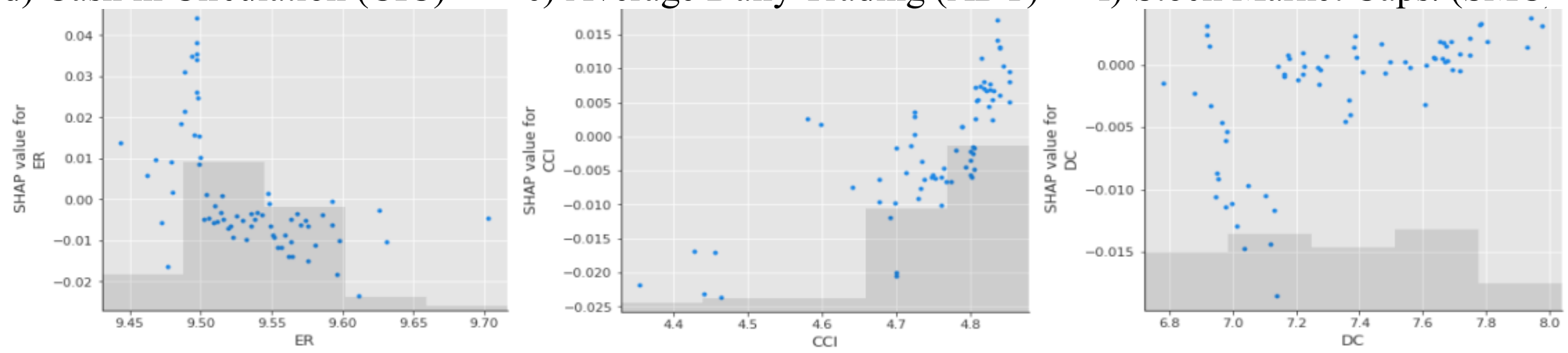

g) Exchange Rate (ER) h) Consumer Conf. Index (CCI) i) Delivery Channel (DC)

**Appendix C:** Functional Form plot of all explanatory variables, Source: Author's calculation

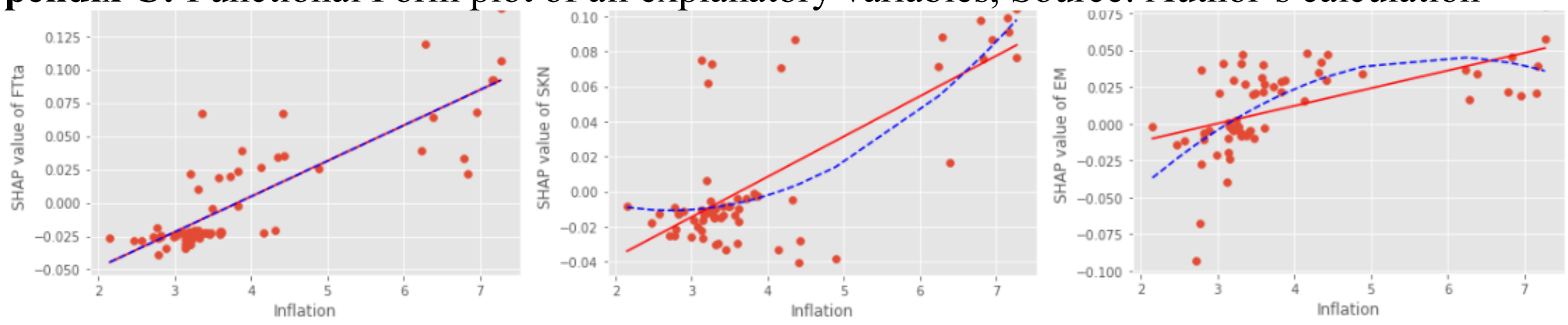

a) Fund Transfer (to abroad) b) National Clearing (SKN) c) Electronic Money

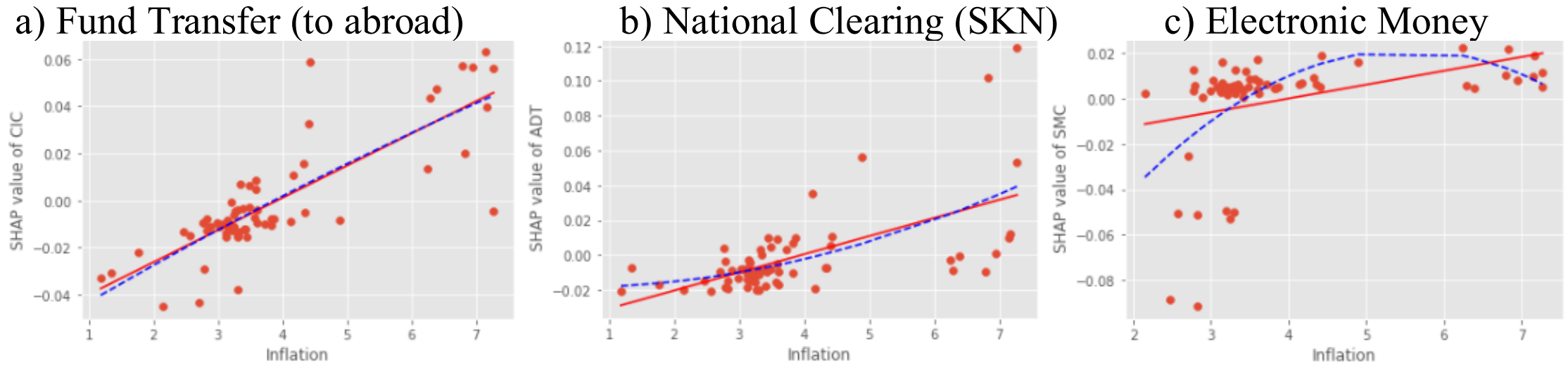

d) Cash in Circulation (CIC) e) Average Daily Trading (ADT) f) Stock Market Caps (SMC)

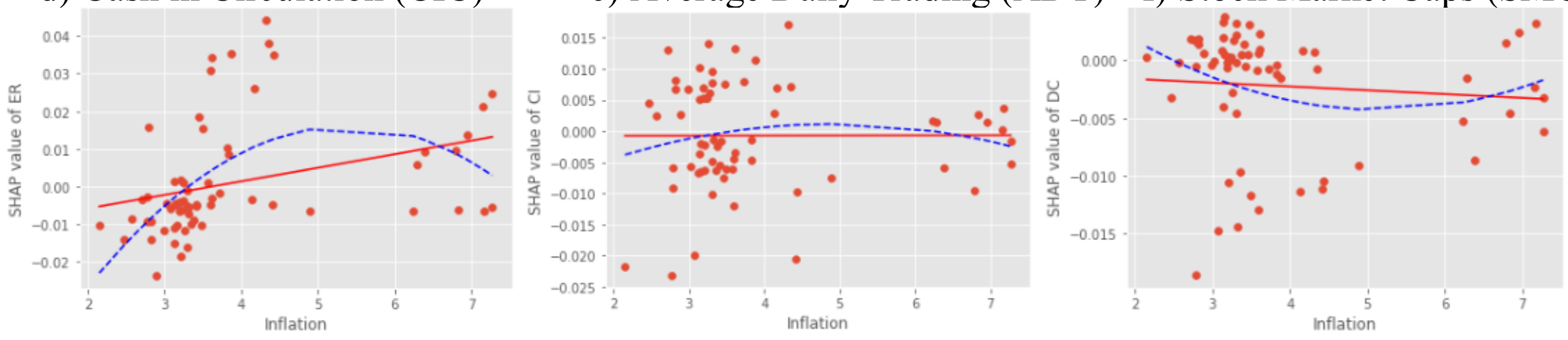

g) Exchange Rate (ER) h) Concumer confidence Index i) Delivery Channel (DC)